\documentclass[twocolumn,letterpaper,aps,prl,longbibliography,superscriptaddress,showpacs,floatfix]{revtex4-1}
\RequirePackage{lineno}
\usepackage{url}
\usepackage{cancel}
\usepackage[colorlinks,linkcolor=blue,citecolor=blue,filecolor=black,urlcolor=blue]{hyperref}
\usepackage[normalem]{ulem} 
\usepackage{epsfig,graphics}
\usepackage{graphicx}
\usepackage{dcolumn}
\usepackage{bm}
\usepackage[usenames]{color}
\usepackage{ulem}
\usepackage{amssymb}
\usepackage{xspace} 
\usepackage{amsmath}
\usepackage{multirow}
\usepackage{float}
\usepackage{harpoon}
\usepackage{MnSymbol}
\usepackage{appendix}
\usepackage{color}


\makeatletter 
\@addtoreset{equation}{section}
\makeatother  

\newcommand{\Yphi}{$Y(\Delta\phi)$}
\newcommand{\YphiTempl}{Y(\Delta\phi)^{\mathrm {templ}}}
\newcommand{\YphiRidge}{Y(\Delta\phi)^{\mathrm{ridge}}}

\newcommand{\Yphipp}{Y(\Delta\phi)^{pp}}

\newcommand{\dphi}{\Delta\phi}
\newcommand{\TopE}{$\sqrt{s_{_{\mathrm{NN}}}}=200$~GeV}

\newcommand{\sqrtsNN}{\mbox{$\sqrt{s_{\mathrm{NN}}}$}}

\newcommand{\pT} {p_{\mathrm{T}}}

\newcommand{\lr}[1]{\left\langle #1\right\rangle}

\newcommand{\Dphi}{\mbox{$\Delta \phi$}}
\newcommand{\Deta}{\mbox{$\Delta \eta$}}
\newcommand{\nch}{N_{\mathrm{ch}}}

\newcommand{\heau}{\mbox{$^{3}$He$+$Au}\xspace}
\newcommand{\dau}{\mbox{$d$$+$Au}\xspace}
\newcommand{\pp}{\mbox{$p$$+$$p$}\xspace}
\newcommand{\pau}{\mbox{$p$$+$Au}\xspace}

\usepackage{CJK}
\hfuzz=\maxdimen
\begin{document}
\title{Measurements of the elliptic and triangular azimuthal anisotropies in central $^{3}$He+Au, $d$+Au and $p$+Au collisions at $\sqrtsNN$ = 200 GeV}

\affiliation{Abilene Christian University, Abilene, Texas   79699}
\affiliation{Alikhanov Institute for Theoretical and Experimental Physics NRC "Kurchatov Institute", Moscow 117218}
\affiliation{Argonne National Laboratory, Argonne, Illinois 60439}
\affiliation{American University of Cairo, New Cairo 11835, New Cairo, Egypt}
\affiliation{Ball State University, Muncie, Indiana, 47306}
\affiliation{Brookhaven National Laboratory, Upton, New York 11973}
\affiliation{University of Calabria \& INFN-Cosenza, Rende 87036, Italy}
\affiliation{University of California, Berkeley, California 94720}
\affiliation{University of California, Davis, California 95616}
\affiliation{University of California, Los Angeles, California 90095}
\affiliation{University of California, Riverside, California 92521}
\affiliation{Central China Normal University, Wuhan, Hubei 430079 }
\affiliation{University of Illinois at Chicago, Chicago, Illinois 60607}
\affiliation{Creighton University, Omaha, Nebraska 68178}
\affiliation{Czech Technical University in Prague, FNSPE, Prague 115 19, Czech Republic}
\affiliation{National Institute of Technology Durgapur, Durgapur - 713209, India}
\affiliation{ELTE E\"otv\"os Lor\'and University, Budapest, Hungary H-1117}
\affiliation{Frankfurt Institute for Advanced Studies FIAS, Frankfurt 60438, Germany}
\affiliation{Fudan University, Shanghai, 200433 }
\affiliation{University of Heidelberg, Heidelberg 69120, Germany }
\affiliation{University of Houston, Houston, Texas 77204}
\affiliation{Huzhou University, Huzhou, Zhejiang  313000}
\affiliation{Indian Institute of Science Education and Research (IISER), Berhampur 760010 , India}
\affiliation{Indian Institute of Science Education and Research (IISER) Tirupati, Tirupati 517507, India}
\affiliation{Indian Institute Technology, Patna, Bihar 801106, India}
\affiliation{Indiana University, Bloomington, Indiana 47408}
\affiliation{Institute of Modern Physics, Chinese Academy of Sciences, Lanzhou, Gansu 730000 }
\affiliation{University of Jammu, Jammu 180001, India}
\affiliation{Joint Institute for Nuclear Research, Dubna 141 980}
\affiliation{Kent State University, Kent, Ohio 44242}
\affiliation{University of Kentucky, Lexington, Kentucky 40506-0055}
\affiliation{Lawrence Berkeley National Laboratory, Berkeley, California 94720}
\affiliation{Lehigh University, Bethlehem, Pennsylvania 18015}
\affiliation{Max-Planck-Institut f\"ur Physik, Munich 80805, Germany}
\affiliation{Michigan State University, East Lansing, Michigan 48824}
\affiliation{National Research Nuclear University MEPhI, Moscow 115409}
\affiliation{National Institute of Science Education and Research, HBNI, Jatni 752050, India}
\affiliation{National Cheng Kung University, Tainan 70101 }
\affiliation{The Ohio State University, Columbus, Ohio 43210}
\affiliation{Panjab University, Chandigarh 160014, India}
\affiliation{NRC "Kurchatov Institute", Institute of High Energy Physics, Protvino 142281}
\affiliation{Purdue University, West Lafayette, Indiana 47907}
\affiliation{Rice University, Houston, Texas 77251}
\affiliation{Rutgers University, Piscataway, New Jersey 08854}
\affiliation{University of Science and Technology of China, Hefei, Anhui 230026}
\affiliation{South China Normal University, Guangzhou, Guangdong 510631}
\affiliation{Sejong University, Seoul, 05006, South Korea}
\affiliation{Shandong University, Qingdao, Shandong 266237}
\affiliation{Shanghai Institute of Applied Physics, Chinese Academy of Sciences, Shanghai 201800}
\affiliation{Southern Connecticut State University, New Haven, Connecticut 06515}
\affiliation{State University of New York, Stony Brook, New York 11794}
\affiliation{Instituto de Alta Investigaci\'on, Universidad de Tarapac\'a, Arica 1000000, Chile}
\affiliation{Temple University, Philadelphia, Pennsylvania 19122}
\affiliation{Texas A\&M University, College Station, Texas 77843}
\affiliation{University of Texas, Austin, Texas 78712}
\affiliation{Tsinghua University, Beijing 100084}
\affiliation{University of Tsukuba, Tsukuba, Ibaraki 305-8571, Japan}
\affiliation{University of Chinese Academy of Sciences, Beijing, 101408}
\affiliation{Valparaiso University, Valparaiso, Indiana 46383}
\affiliation{Variable Energy Cyclotron Centre, Kolkata 700064, India}
\affiliation{Wayne State University, Detroit, Michigan 48201}
\affiliation{Yale University, New Haven, Connecticut 06520}

\author{M.~I.~Abdulhamid}\affiliation{American University of Cairo, New Cairo 11835, New Cairo, Egypt}
\author{B.~E.~Aboona}\affiliation{Texas A\&M University, College Station, Texas 77843}
\author{J.~Adam}\affiliation{Czech Technical University in Prague, FNSPE, Prague 115 19, Czech Republic}
\author{J.~R.~Adams}\affiliation{The Ohio State University, Columbus, Ohio 43210}
\author{G.~Agakishiev}\affiliation{Joint Institute for Nuclear Research, Dubna 141 980}
\author{I.~Aggarwal}\affiliation{Panjab University, Chandigarh 160014, India}
\author{M.~M.~Aggarwal}\affiliation{Panjab University, Chandigarh 160014, India}
\author{Z.~Ahammed}\affiliation{Variable Energy Cyclotron Centre, Kolkata 700064, India}
\author{A.~Aitbaev}\affiliation{Joint Institute for Nuclear Research, Dubna 141 980}
\author{I.~Alekseev}\affiliation{Alikhanov Institute for Theoretical and Experimental Physics NRC "Kurchatov Institute", Moscow 117218}\affiliation{National Research Nuclear University MEPhI, Moscow 115409}
% ...

\collaboration{STAR Collaboration}\noaffiliation

\begin{abstract}

The elliptic ($v_2$) and triangular ($v_3$) azimuthal anisotropy coefficients in central $^{3}$He+Au, $d$+Au, and $p$+Au collisions at $\sqrtsNN$ = 200 GeV are measured as a function of transverse momentum ($\pT$) at mid-rapidity ($|\eta|<$0.9), via the azimuthal angular correlation between two particles both at $|\eta|<$0.9. While the $v_2(\pT)$ values depend on the colliding systems, the $v_3(\pT)$ values are system-independent within the uncertainties, suggesting an influence on eccentricity from sub-nucleonic fluctuations in these small-sized systems. These results also provide stringent constraints for the hydrodynamic modeling of these systems. 

\end{abstract}

\maketitle
Relativistic heavy-ion collisions produce the Quark Gluon Plasma (QGP), which has an anisotropic transverse energy density profile~\cite{Arsene:2004fa,Adcox:2004mh,Back:2004je,Adams:2005dq,ROLAND201470}. The eccentricity of this density profile can induce anisotropic pressure gradients, giving rise to strong anisotropies of particle distribution relative to the flow planes $\Psi_n$~\cite{Voloshin:1994mz,Poskanzer:1998yz,Qiu:2011iv}. 
This anisotropy is often quantified via Fourier decomposition of the two-particle correlations in relative azimuthal angle 
$\Delta\phi=\phi_{\alpha}-\phi_{\beta}$~\cite{Poskanzer:1998yz,Lacey:2005qq} for the particles $\alpha$ and $\beta$ as a function of transverse momentum ($\pT$):

\begin{equation}
\begin{split}
\label{eq:2}
\frac{dN^{\mathrm{pairs}}}{d\Delta\phi}&\propto1+2\sum_{n=1}^{\infty}c_{n}\cos(n\Delta\phi), \\
c_{n}(\pT^{\alpha},\pT^{\beta})  &= {v_n(\pT^{\alpha})v_n(\pT^{\beta})+ \delta_{\mathrm{NF}}},
\end{split}
\end{equation}
where $\delta_{\text{NF}}$ represents the correlation unrelated to collective effects (``nonflow" correlation). The $v_{2}\{2\}$ and $v_{3}\{2\}$ (termed $v_2$ and $v_3$ ) harmonics that are linearly related to the respective eccentricities of initial energy density spatial distribution,  
$\varepsilon_{2}\{2\}$ and $\varepsilon_{3}\{2\}$, provide an important model constraint on the specific shear viscosity of the QGP produced in large- to moderate-sized A+A systems such as Pb+Pb, Au+Au and Cu+Cu collisions~\cite{Song:2010mg, Niemi:2012aj,Gardim:2014tya, Fu:2015wba,Holopainen:2010gz,Qin:2010pf,Qiu:2011iv,Gale:2012rq,Liu:2018hjh,Schenke:2011tv}.

For small-sized systems such as $p$+$p$, $p$/$d$/$^{3}$He+A collisions, the azimuthal anisotropies have been extensively measured at RHIC \cite{PHENIX:2014fnc, PHENIX:2015idk,PHENIX:2018lia,PHENIX:2021ubk,PHENIX:2022nht,2015265,STAR:2019zaf} and the LHC \cite{Chatrchyan:2013nka,Abelev:2012ola,Aad:2012gla,Aaboud:2017acw}. Numerical simulations suggest that hydrodynamics remains applicable even when the system size is of the order of the inverse temperature~\cite{Chesler:2016ceu}. However, the influence of sub-nucleonic fluctuations on the initial geometry, which is negligible for larger-sized systems, has not been charted for small-sized systems. Such fluctuations can result from a spatially inhomogeneous gluon field distribution inside the nucleon~\cite{PhysRevC.94.024919,PhysRevLett.108.252301}. Table~\ref{tab:eccentricity} gives an illustrative comparison of the eccentricities for \heau, \dau, and \pau collisions from four scenarios, all based on Glauber models and labeled as $a$, $b$, $c$, and $d$. Model $a$ corresponds to the mean eccentricities reported in Ref. ~\cite{PhysRevLett.113.112301}; it uses the default Glauber model to calculate the nucleon position and does not have quantum fluctuations. Model $b$ also uses the default Glauber for nucleon position but includes quantum fluctuations characterized by a smoothly distributed Gaussian-like gluon field inside each nucleon~\cite{PhysRevC.94.024919}. In Models $c$ and $d$, there are several gluon fields surrounding the valence quarks inside the nucleon instead of one gluon field as in Model $b$. The distribution of the gluon field is Gaussian-like in Model $c$~\cite{PhysRevC.94.024919} but is lumpy for the IP-Glasma framework~\cite{PhysRevLett.108.252301,PHENIX:2021ubk} used in Model $d$. Table~\ref{tab:eccentricity} shows that the system dependence of $\varepsilon_{2,3}$ is strongly influenced by sub-nucleonic fluctuations, suggesting that measurements of the system dependence of $v_{2,3}(\pT)$ can provide invaluable constraints on the role of such fluctuations in small-sized systems and give insights into the structure of the nucleon.

Furthermore, the anisotropy may also originate from non hydrodynamic modes~\cite{Bzdak:2014dia,Zhou:2015iba,Romatschke:2015dha,Bierlich:2017vhg,Mark:2018,Dusling:2013oia,PhysRevC.100.064905} and/or large hydrodynamic gradient-expansion corrections~\cite{Denicol:2012cn,Florkowski:2016zsi} due to the short lifetime of the created medium. Therefore, whether hydrodynamics can extend its success from large- and moderate-sized systems to small-sized systems remains uncertain.

\begin{table}[t]
\caption{
Comparison of the system dependence of $\varepsilon_{2}$($\varepsilon_{3}$) in central \heau, \dau, and \pau collisions from four Glauber-based models (see text). For Model a and d, the $\left<\varepsilon_{2}\right>$ and $\left<\varepsilon_{3}\right>$ values are obtained for impact parameter b $< 2$~fm; For Model b and c, the $\varepsilon_{n}$ values are obtained as $\sqrt{\left<\varepsilon_{n}^2\right>}$ for $0-10\%$ \heau and \dau, and $0-2\%$ \pau collisions selected by multiplicity. The relative difference of the $\varepsilon_n$ values for the three systems is not strongly influenced by the difference in event selection nor the $\varepsilon_{n}$ definition. The statistical uncertainties are much less than 1\%.}
\begin{ruledtabular} \begin{tabular}{lcccc}
  Model &  a~\cite{PhysRevLett.113.112301,sonic} &  b~\cite{PhysRevC.94.024919}  &  c~\cite{PhysRevC.94.024919} & d~\cite{PhysRevLett.108.252301,PHENIX:2021ubk}\\
   & $\varepsilon^{a}_{2}$($\varepsilon^{a}_{3}$) &
   $\varepsilon^{b}_{2}$($\varepsilon^{b}_{3}$) & $\varepsilon^{c}_{2}$($\varepsilon^{c}_{3}$) &
   $\varepsilon^{d}_{2}$($\varepsilon^{d}_{3}$)\\
\hline
    $^{3}$He+Au  & 0.50(0.28) & 0.52(0.35) & 0.53(0.38) & 0.64(0.46)\\
    $d$+Au  & 0.54(0.18) & 0.51(0.32) & 0.53(0.36) & 0.73(0.40)\\
    $p$+Au  & 0.23(0.16) & 0.34(0.27) & 0.41(0.34) & 0.50(0.32) \\
\end{tabular} \end{ruledtabular}
\label{tab:eccentricity}
\end{table}

Prior measurements of $v_{2,3}(\pT)$ for \heau, \dau, and \pau collisions have been reported by the PHENIX collaboration~\cite{PHENIX:2018lia,PHENIX:2021ubk,PHENIX:2022nht}. These measurements, which utilized correlations between particles at middle and backward pseudorapidity ($\eta$), indicated values compatible with the system dependence of $\varepsilon^{a}_{n}$ and little influence from sub-nucleonic fluctuations. Here, we present complementary $v_{n}$ measurements for pseudorapidity $|\eta| <$ 0.9 via correlations between particles both at middle pseudorapidity to investigate further a possible role for sub-nucleonic fluctuations. The two-particle azimuthal correlations employed for the measurements, suppress the influence of nonflow correlations via the requirement $|\Delta\eta| >1.0$ in conjunction with three established methods of nonflow subtraction~\cite{adams:2004ja,Adams2005:aj,2017193,2015333,PhysRevC.90.044906,PhysRevC.98.014912,Aad:2016}.

The \heau, \dau, \pau, and \pp data used in this analysis are collected with a minimum bias (MB) and a high multiplicity (HM) triggers in 2014, 2015, and 2016 experimental runs of the STAR experiment at $\sqrtsNN$ = 200 GeV. Events were selected to be within a radius $r<2$~cm relative to the beam axis and within specific ranges of the center of the TPC in the direction along the beam axis, $v_{z}$ with the values $\pm$ 30~cm for \heau,  $\pm$ 15~cm for \dau,  $\pm$ 20~cm for \pau and $\pm$ 20~cm for \pp. The MB trigger for \pp, \pau, and \dau collisions required a coincidence between both sides of the Vertex Position Detectors (VPD)~\cite{STARVPD} along the beam pipe, which span the range 4.4 $< |\eta| <$ 4.9. The MB trigger for \heau employed a coincidence between both sides of the VPD, a coincidence between both sides of the Beam-Beam Counters (BBC)~\cite{STARtrig} which span the range $3.3 < |\eta| < 5.1$, and a neutron hit in the Zero Degree Calorimeter (ZDC)~\cite{STARZDC} on the Au-going side. For \pau collisions, the MB triggers were augmented with a number-of-hits cut of more than 80 in the Barrel Time of Flight (BTOF) detector with $|\eta| <1$~\cite{STARTOF} to obtain the HM triggers.

The collision centrality is determined via Monte Carlo Glauber model calculations~\cite{STARGlauber,Loizides_2015} tuned to match the distribution of the number of reconstructed charged tracks before efficiency correction ($\nch^{\mathrm{off}}$) in the MB events. To count $\nch^{\mathrm{off}}$, tracks are selected to have $|\eta| <$ 0.9 and 0.2 $< \pT <$ 3.0 GeV/c with a matched hit in the BTOF detector. In this work, we use the top $0-10\%$ centrality for $d$+Au,  and both $0-10\%$ and $10-20\%$ for $^{3}$He+Au collisions. For \pau collisions, the HM datasets, supplemented with a threshold cut on $\nch^{\mathrm{off}}$, are used to select ultra-central (UC) events. This choice facilitates the comparison of the $v_{n}$ measurements for UC \pau, $0-10\%$ \dau and $10-20\%$ \heau with comparable track multiplicity after efficiency correction($\lr{\nch}$), as listed in Table~\ref{t:nch}. Note that $\lr{\nch}$ for the UC \pau is also similar to that for the $0-2\%$ \pau MB data sample.  The charged-hadron efficiency is obtained via the embedding of simulated charged pions~\cite{GEANT, FINE200176} into actual data. The systematic uncertainties for $\lr{\nch}$ listed in Table~\ref{t:nch} arise mainly from the uncertainties of $\pi^{\pm}$ reconstruction efficiency. There are additional 10\% overall systematic uncertainties that arise from the efficiency estimations, which combine $\pi^{\pm}$, $K^{\pm}$, and (anti-)protons together. And such uncertainties are largely canceled out in flow measurements.

\begin{table}[htbp]
\caption{\label{t:nch}The average of efficiency-corrected multiplicity, $\left\langle \nch \right\rangle $, in MB \pp and central $p$/$d$/\heau collisions at \TopE. The uncertainties reflect  both systematic and statistical uncertainties.}
\begin{ruledtabular}
\begin{tabular}{cccccc}
     & MB & UC & $0-10\%$ & $10-20\%$ & $0-10\%$ \\
     &$p+p$ & \pau & \dau & \heau & \heau\\
     \hline
     $\lr{\nch}$ & 4.7$\pm$0.3 & 34.1$\pm$1.7 & 35.6$\pm$1.8 & 33.1$\pm$1.7 & 47.7$\pm$2.4 \\
\end{tabular}
\end{ruledtabular}
\end{table}

The charged particles detected in the Time Projection Chamber(TPC)~\cite{STARTPC} are used to construct two-particle yield distributions \Yphi\  = $1/N_{\mathrm{Trig}}dN/d\Delta\phi$ with efficiency correction applied. The detector acceptance effects have been corrected by pairs from different events. The effect of multiple collisions from a bunch crossing (pile-up) is primarily suppressed by requiring a matched hit in the BTOF detector or one of the two layers of silicon strip sensors of the Heavy Flavor Tracker (HFT) detector~\cite{STARHFT}, both of which have fast responses.

\begin{figure}[ht]
  \includegraphics[width=1.0\linewidth]{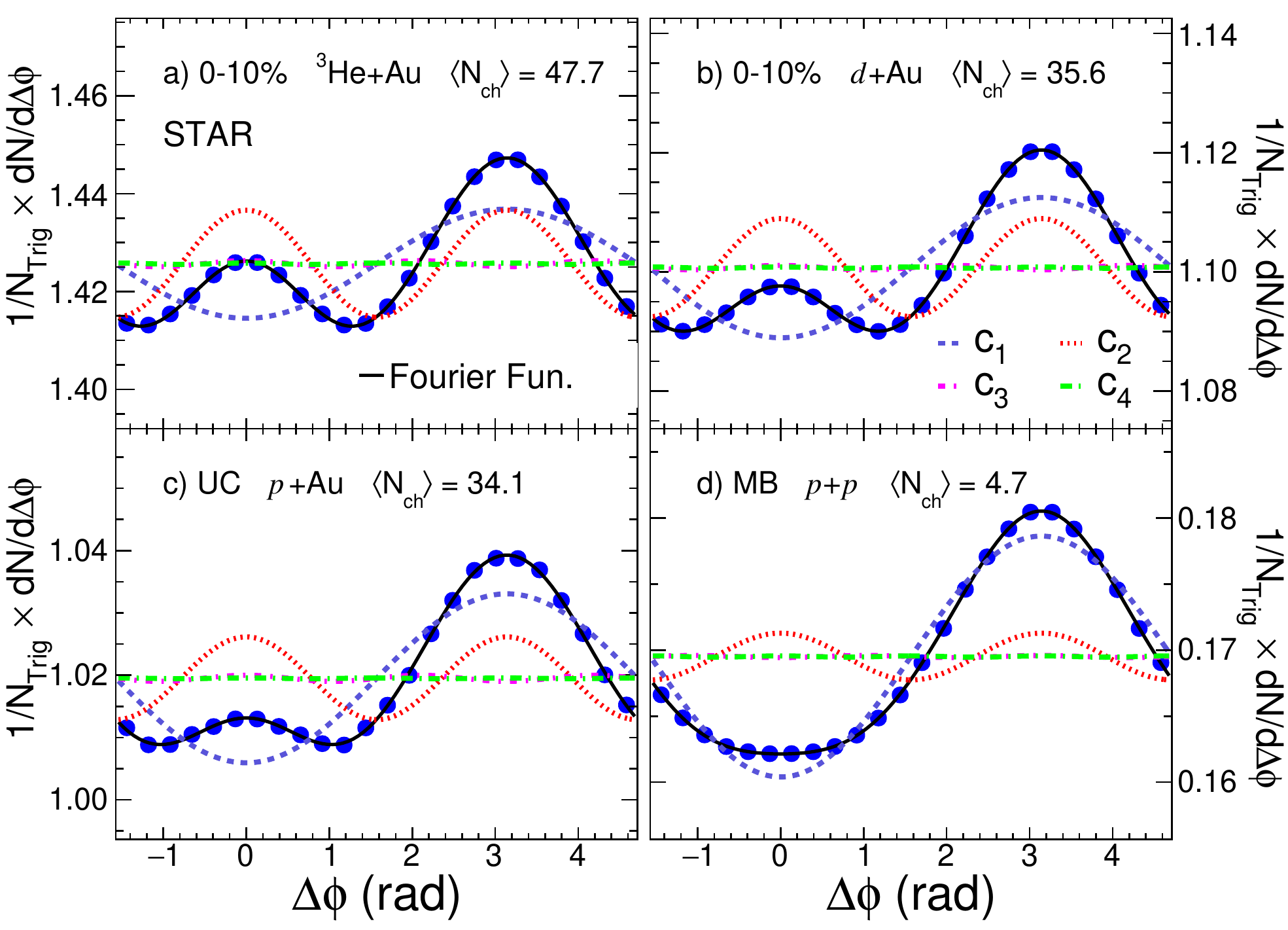}
  \caption{
	Two-particle per-trigger yield distributions for \heau, \dau, \pau, and \pp collisions at $\sqrtsNN$ = 200 GeV as indicated. The trigger and associated particles are selected in the range $0.2 < \pT < 2.0$~GeV/c and 1.0 $< |\Deta| <$ 1.8. An illustration of the Fourier functions fitting procedure to estimate the nonflow contributions and extract the $v_{2,3}$ is also shown.
}
 \label{fig:corr_functions}
\end{figure}

Figure~\ref{fig:corr_functions}(a)-(d) show the distributions \Yphi\ for central \heau, \dau, \pau, and MB \pp collisions as a function of $\dphi$. The trigger ($\mathrm{Trig.}$)- and the associated (Assoc.)-particles are measured in the range $0.2 < \pT < 2.0$~GeV/$c$ and 1.0 $< |\Deta| <$ 1.8. The near-($|\Dphi| <$ 1.0) and away-side($|\Dphi-\pi| <$ 1.0) distributions for $^{3}$He+Au, $d$+Au and $p$+Au indicate a sizable impact from nonflow correlations that can be removed with three subtraction methods (termed I, II, III) that utilize the correlation functions from MB $\pp$ as outlined below. Note the similarity between the away-side distributions for \heau, \dau, \pau,  and that for $\pp$, which is dominated by nonflow.

In all methods, a Fourier function fit of the measured \Yphi\ distributions is employed to extract $v_{n}(\pT^{\mathrm{Trig.}})$:
\begin{eqnarray}
 Y(\Delta\phi, \pT^{\mathrm{Trig.}}) = c_{0}(1 + \sum_{n=1}^{4} \, 2c_{n}\, \cos ( n \,\Delta\phi )). 
 \label{eq:fourier}
\end{eqnarray}

\noindent The non-flow contributions are subtracted with

\begin{eqnarray}
c_{n}^{\mathrm{sub}} = c_{n} - c^{\mathrm{nonflow}}_{n} = c_{n} - c_{n}^{pp}\times f
\label{eq:c2_collective_noncollective}
\end{eqnarray}
where the $c_{n}^{\mathrm{sub}}$ is $c_{n}$ after nonflow subtraction. The methods differ from each other in how the scale factor $f$ is estimated. The $c_n$ is simply the product of $v_n$ for trigger- and associated-particles, i.e. $c_{n}=v_{n}^{\mathrm{Trig.}} \times v_{n}^{\mathrm{Assoc.}}$

Method I assumes that the nonflow correlations between \pp and $p$/$d$/$^3$He+Au are the same. Thus the factor $f$ is equal to the ratio of the integral yield of \Yphi\ ($c_0$) due to the multiplicity dilution. Then $f = {c_{0}^{pp}}/{c_{0}}$. This method is found to be similar to the so-called ``scalar product method''~\cite{PhysRevC.64.054901,adams:2004ja, Adams2005:aj} from testing.

The nonflow contributions in \pp collisions could be different from those in $p$/$d$/$^3$He+Au collisions; such differences are corrected in Methods II and III by looking into the near-side yield and away-side shape of the nonflow correlations.

Method II estimates the nonflow contribution to the near-side yield ($Y^{N}$) from the difference between the \Yphi\ yield measured for 0.2 $< |\Delta\eta| < $ 0.5 and 1.0 $< |\Delta\eta| <$ 1.8, as outlined in Refs.~\cite{2017193,2015333,PhysRevC.90.044906}. Then
$f = ({Y^{N}}/{Y^{N}_{pp}})\times({c_{0}^{pp}}/{c_{0}})$.

With the $|\Deta|>$1.0 requirement, the residual nonflow arises primarily from the away-side correlations, which is dominated by the $c_1$ component. The Method III uses $c_1$ to estimate $f$ directly~\cite{PhysRevC.98.014912}, then $f = {c_{1}}/{c_{1}^{pp}}$.

Method III is also similar to the Template fit method~\cite{Aad:2016} as shown in the supplemental document.

Since $v_{n}^{\mathrm{Assoc.}}\equiv\sqrt{c_n}$ for trigger and associated particles in the same $\pT$ range, one has $v_{n}^{\mathrm{Trig.}}=c_{n}/v_{n}^{\mathrm{Assoc.}}$. Similarly, the $v_n$ after nonflow subtraction ($v_{n}^{\mathrm{sub}}$) is 
computed as $v_{n}^{\mathrm{sub, Trig.}}=c_{n}^{\mathrm{sub}}/v_{n}^{\mathrm{sub, Assoc.}}$. 
\begin{figure}[]
\begin{center}
  \includegraphics[width=1.0\linewidth]{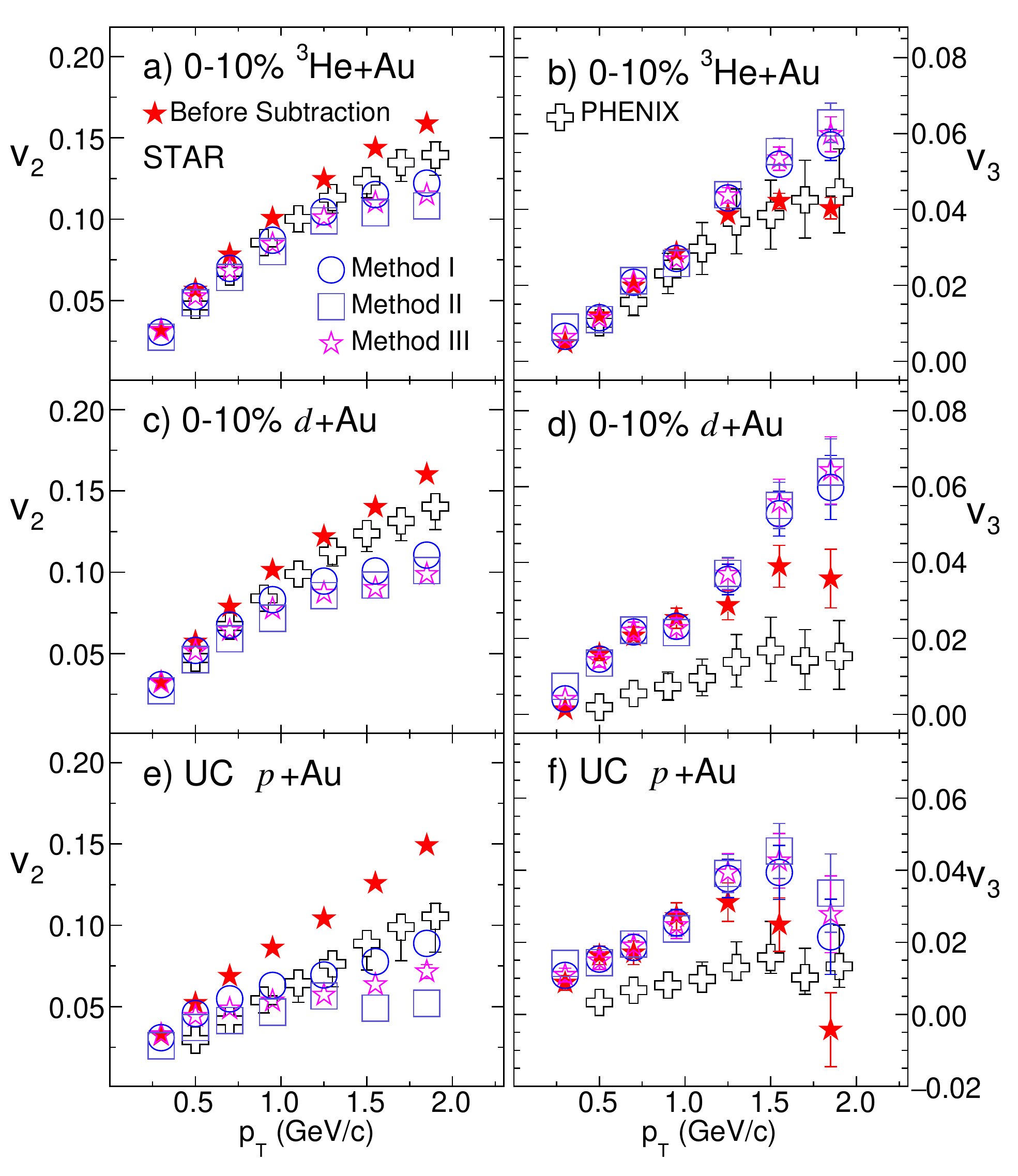}
  \caption{
Comparison of the $v_2$ (left column) and $v_3$ (right column) in $0\%-10\%$ $^{3}$He+Au, $0\%-10\%$ $d$+Au, and UC $p$+Au collisions before and after three different nonflow subtraction methods (see text). Only statistical uncertainties are shown. The PHENIX measurements with statistical and systematic uncertainties are also shown.
}
  \label{fig:method}
  \end{center}
\end{figure}
%

The systematic uncertainties associated with $v_{2,3}(\pT)$ have four main contributions: (i) variation of associated detectors used in track matching, (ii) background tracks, (iii) residual pile-up effects, and (iv) uncertainties for nonflow subtraction. (i)~A comparison of the results obtained with TOF matching and HFT matching shows a difference in $v_{2}$($v_{3}$) of less than 3\%(10\%) for all three systems. (ii)~The track background uncertainty is estimated by varying the cut on the number of TPC space points used for track reconstruction from 15 to 25. The resulting values vary less than 5\%(10\%) in $v_{2}$($v_{3}$). (iii)~The impact of residual pileup is estimated by comparing results obtained from data with different beam luminosities, giving a difference of less than 2\%(5\%) for $v_{2}$($v_{3}$) for all three systems. (iv)~
The uncertainties associated with the nonflow subtraction is estimated by comparing between subtraction methods and $\Delta\eta$ cuts ($|\Delta\eta|>$ 0.8, 1.2 and 1.4), as well as between the same-charge and opposite-charge particle pairs. The results from Method III, which are close to the average of the results from the three methods, are taken as the default, and the differences from the other two methods and variations are taken as the systematic uncertainties. The resulting uncertainty is up to 25\%(30\%) in $v_{2}$($v_{3}$). A study based on the HIJING model~\cite{HIJING} (shown in the supplemental documents) indicates that the uncertainties for nonflow subtraction are within the systematic uncertainties assigned here. 

Figure~\ref{fig:method} shows a comparison of the $v_{n}$ values extracted for central $^{3}$He+Au, $d$+Au, and $p$+Au collisions before and after nonflow subtraction. The away-side nonflow correlations give a positive contribution to $v_2$ and a negative one to $v_3$. Therefore, the subtraction decreases the magnitude of $v_2$ as shown in the left panels of Fig.~\ref{fig:method}, but increases the magnitude of $v_3$ as shown in the right panels. The comparison also indicates that the respective methods give similar results after subtraction. 

Comparisons to the published PHENIX measurements \cite{PHENIX:2018lia,PHENIX:2021ubk} indicate that, within the uncertainties, the $v_2(\pT)$ results for all three collision systems and the $v_3(\pT)$ results for $^3$He+Au collisions from both experiments are in reasonable agreement with a maximum difference $\approx 25\%$. However, the STAR $v_3(\pT)$ measurements for $p$+Au and $d$+Au collisions are about a factor of 3 larger than those reported by PHENIX. This difference is insensitive to the different centrality definitions employed in the two experiments (see supplemental information). The root cause of this discrepancy is still not fully understood. On the other hand, a recent model study~\cite{Zhao:2022ugy} indicates that up to 50\% of this $v_3(\pT)$ discrepancy could result from the larger longitudinal de-correlation possible in the PHENIX measurements. 
However, calculations from this model systematically under-predict the individual STAR and PHENIX $v_3(\pT)$ measurements in \pau collisions. The data-model comparison may improve in the future with the inclusion of effects such as nonflow and pre-hydrodynamic flow effects in the calculations.

\begin{figure}[]
\begin{center}
  \includegraphics[width=1.0\linewidth]{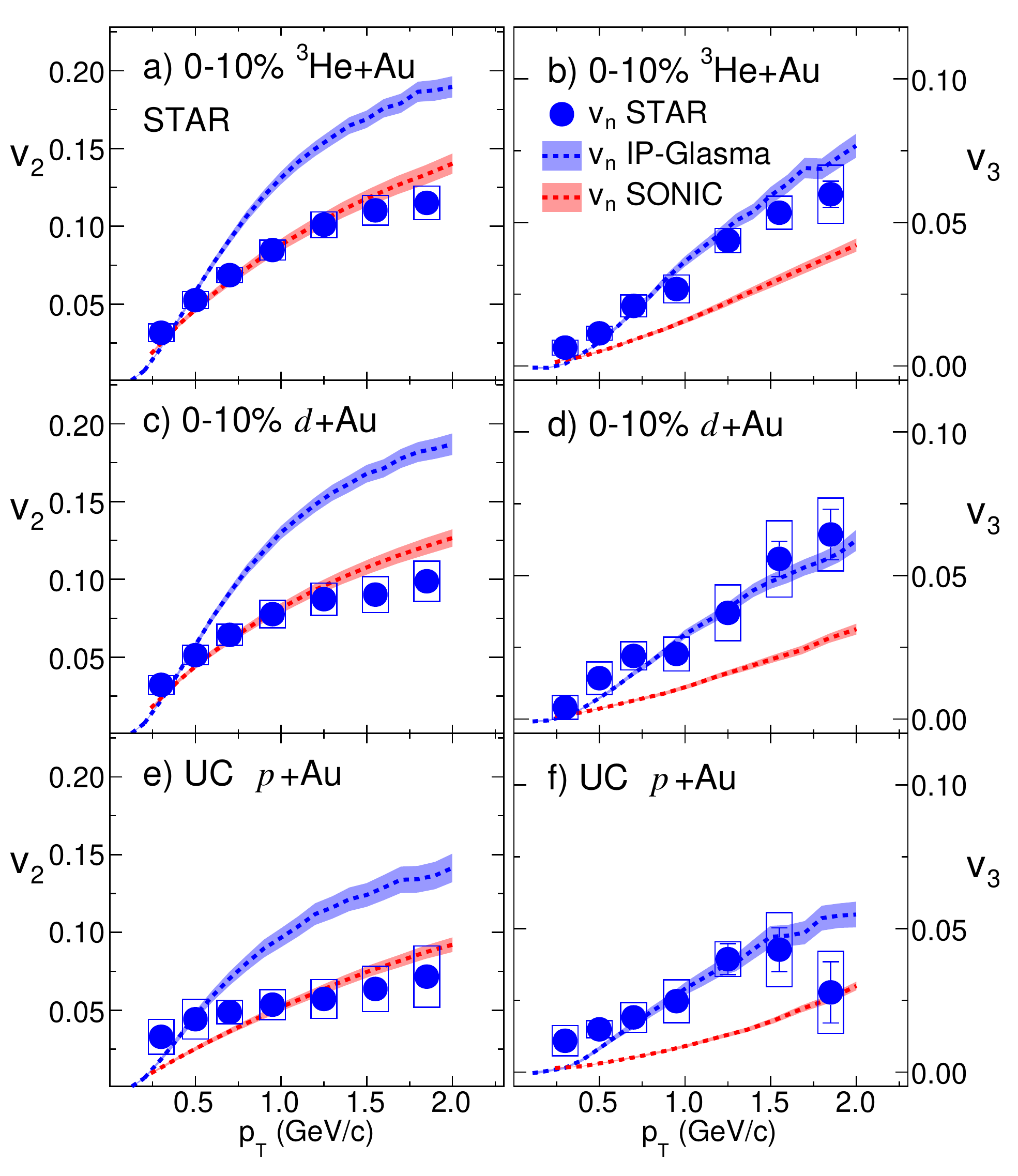}
  \caption{
	Comparison of the $v_{2,3}$ from data and hydrodynamic model calculations in $0-10\%$ $^{3}$He+Au, $0-10\%$ $d$+Au, and UC $p$+Au collisions. The theory curves are obtained from the {\sc sonic}~\cite{PhysRevLett.113.112301,sonic} and the IP-Glasma+MUSIC ~\cite{Schenke_2020,schenke2020running} hydrodynamic models.
}
  \label{fig:v2v3theo}
  \end{center}
\end{figure}

We compare our results to two hydrodynamic model calculations- {\sc sonic}~\cite{PhysRevLett.113.112301,sonic} and IP-Glasma+MUSIC~\cite{Schenke_2020,schenke2020running}  - in Fig.~\ref{fig:v2v3theo}. The pre-existing calculations from SONIC are only available for the $0-5\%$ centrality, but the differences from the centrality mismatch are expected to be around 10\%. The {\sc sonic} model, which roughly describes the PHENIX measurements \cite{PHENIX:2018lia}, employs initial eccentricity from nucleon Glauber without sub-nucleonic fluctuations (Model $a$). The SONIC calculations show reasonable agreement with the current measurements for $v_2(\pT)$ but under-estimate the $v_3(\pT)$ in \heau and significantly under-estimate the $v_3(\pT)$ in \dau and \pau collisions by more than 100\% . This under-prediction could be due to the much smaller $\varepsilon_3$ values without sub-nucleonic fluctuations employed in the calculations. Interestingly, the SONIC calculations give a reasonable prediction of $v_2(\pT)$ for \pau with the much smaller $\varepsilon_2$ value indicated in Table I. It is currently unclear if this is related to possible uncertainties in the hydrodynamic gradient-expansion corrections or other sources.

The IP-Glasma+MUSIC model includes sub-nucleonic fluctuations, momentum correlations, and pre-hydrodynamic flow in the initial state. For the final state, it includes  viscous hydrodynamic evolution, and the UrQMD model for evolution in the hadronic phase~\cite{Schenke_2020,schenke2020running}. It is tuned to describe the data for large-sized systems and then extrapolated to small-sized systems without further tuning. In contrast to the {\sc sonic} model, the calculations from the IP-Glasma+MUSIC model over-predict the $v_2(\pT)$ data, but show good agreement with the $v_3(\pT)$ data for all three systems. The over-prediction could result from: (i) an overestimate of the system-dependent $\varepsilon_2$ values employed in the calculations (see  Model $d$ in Table~\ref{tab:eccentricity});  (ii) the sizable pre-hydrodynamic flow included in the IP-Glasma+MUSIC model framework.  

Figure~\ref{fig:v2v3theo} shows that both models fail to give a simultaneous description of $v_2(\pT)$ and $v_3(\pT)$, indicating that further studies are required to identify model parameters that regulate the influence of the sub-nucleonic fluctuations on $\varepsilon_{2,3}$, and a possible influence from longitudinal flow de-correlation~\cite{Zhao:2022ugy}.
\begin{figure}[]
  \includegraphics[width=1.0\linewidth]{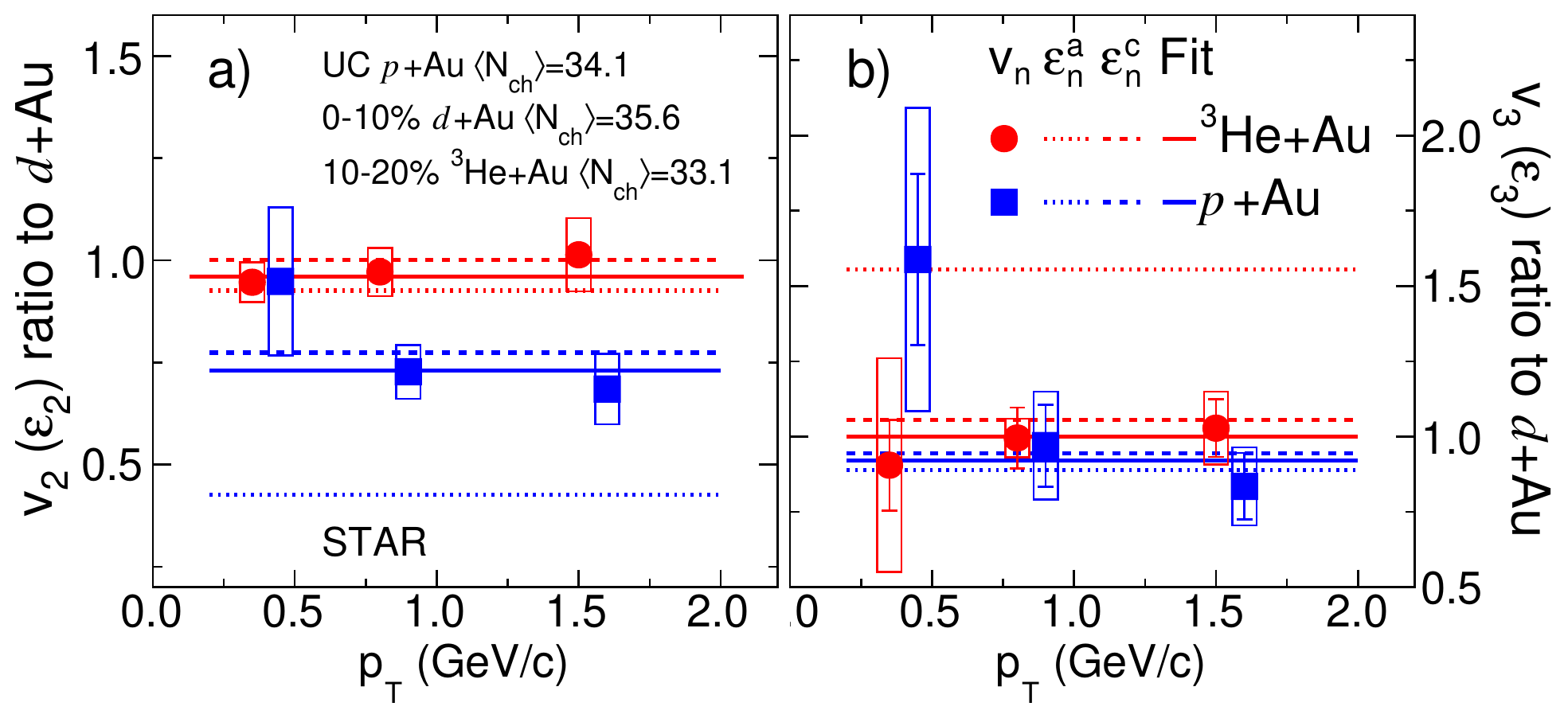}
	\caption{
	Comparison of the ratios of $v_{2}$ (panel a) and $v_{3}$ (panel b) between a given small system and $d$+Au at similar $\lr{\nch}$ for several $\pT$ selections. The solid lines indicate a fit to the data points, and the dashed lines indicate the corresponding eccentricity ratios obtained from Glauber-based model calculations with ($\varepsilon^{c}$, large dash line)~\cite{PhysRevC.94.024919,PhysRevLett.108.252301,PHENIX:2021ubk} and without ($\varepsilon^{a}$, small dashline)~\cite{PhysRevLett.113.112301} sub-nucleonic fluctuations, respectively. 
}
  \label{fig:ratio}
\end{figure}

We further compare the difference between these three systems via $v_n$ ratios at similar mean multiplicity $\lr{\nch}$, as shown in Fig.~\ref{fig:ratio}. Such ratios can give insight into the influence of the initial stage of the collisions since the differences in the final state contributions are expected to be largely canceled for similar multiplicity $\lr{\nch}$~\cite{Liu:2018xae,Liu:2018hjh}. We also compare the $v_{n}$ ratios with the corresponding $\varepsilon_n$ ratios in Fig.~\ref{fig:ratio}; in the absence of other initial state influences, $v_{n}$ is expected to be proportional to $\varepsilon_n$. Hence, the comparison of their ratios can serve as a baseline. The ratio $v_{2,p\mathrm{Au}}/v_{2,d\mathrm{Au}}$ equals to $0.73\pm 0.05$(stat.+syst) from fitting to a constant. It is close to the ratios of $\varepsilon_2$ for the models with sub-nucleonic fluctuations ($\varepsilon^{b,c,d}_{2,p\mathrm{Au}}/\varepsilon^{b,c,d}_{2,d\mathrm{Au}} =$0.65, 0.77 and 0.68, respectively and only model c is shown in Fig.~\ref{fig:ratio}). However, it is 6.0 $\sigma$ away from the ratio $\varepsilon^{a}_{2,p\mathrm{Au}}/\varepsilon^{a}_{2,d\mathrm{Au}} =0.43$ without sub-nucleonic fluctuations. The ratio $v_{3,^3{\mathrm He}+\mathrm{Au}}/v_{3,d\mathrm{Au}} = 1.00\pm 0.09$ is also similar to those for $\varepsilon_3$ from the models with sub-nucleonic fluctuations ( $\varepsilon^{b,c,d}_{3,^3\mathrm{He+Au}}/\varepsilon^{b,c,d}_{3,d\mathrm{Au}}=$1.09, 1.05 and 1.15 respectively). By contrast, it is 6.2 $\sigma$ away from the $\varepsilon^{a}_{3,^3\mathrm{He+Au}}/\varepsilon^{a}_{3,d\mathrm{Au}}=1.56$ (without fluctuations). The comparison suggests that sub-nucleonic fluctuations play a crucial role in establishing the initial state geometry. However, these small systems require further model comparisons to their ratios to ascertain a possible influence from other initial stage contributions, such as pre-hydrodynamics flow.

In summary, we measured $v_{2,3}(\pT)$ in central $^{3}$He+Au, $d$+Au, and $p$+Au collisions at $\sqrtsNN$ = 200 GeV, extracted from two-particle azimuthal angular correlations ($|\Delta\eta|>$1.0) with three subtraction methods designed to mitigate the influence of the nonflow correlations. Results from these methods are consistent within uncertainties. The magnitude of $v_2$ in \pau collisions is lower than that of \dau and \heau collisions, while the magnitude of $v_3$ is system-independent. The measurements are consistent with a significant influence from sub-nucleonic eccentricity fluctuations. Hydrodynamic model comparisons to the data suggest that further model constraints, especially for the theoretical parameters which regulate the sub-nucleonic fluctuations, are required for more detailed characterizations of the azimuthal anisotropy in small-sized systems.

\section*{Acknowledgments}
\begin{acknowledgements}
We thank the RHIC Operations Group and RCF at BNL, the NERSC Center at LBNL, and the Open Science Grid consortium for providing resources and support.  This work was supported in part by the Office of Nuclear Physics within the U.S. DOE Office of Science, the U.S. National Science Foundation, National Natural Science Foundation of China, Chinese Academy of Science, the Ministry of Science and Technology of China and the Chinese Ministry of Education, the Higher Education Sprout Project by Ministry of Education at NCKU, the National Research Foundation of Korea, Czech Science Foundation and Ministry of Education, Youth and Sports of the Czech Republic, Hungarian National Research, Development and Innovation Office, New National Excellency Programme of the Hungarian Ministry of Human Capacities, Department of Atomic Energy and Department of Science and Technology of the Government of India, the National Science Centre and WUT ID-UB of Poland, the Ministry of Science, Education and Sports of the Republic of Croatia, German Bundesministerium f\"ur Bildung, Wissenschaft, Forschung and Technologie (BMBF), Helmholtz Association, Ministry of Education, Culture, Sports, Science, and Technology (MEXT) and Japan Society for the Promotion of Science (JSPS).
\end{acknowledgements}
\bibliography{main-arXiv}{}

\begin{thebibliography}{66}%
\makeatletter
\providecommand \@ifxundefined [1]{%
 \@ifx{#1\undefined}
}%
\providecommand \@ifnum [1]{%
 \ifnum #1\expandafter \@firstoftwo
 \else \expandafter \@secondoftwo
 \fi
}%
\providecommand \@ifx [1]{%
 \ifx #1\expandafter \@firstoftwo
 \else \expandafter \@secondoftwo
 \fi
}%
\providecommand \natexlab [1]{#1}%
\providecommand \enquote  [1]{``#1''}%
\providecommand \bibnamefont  [1]{#1}%
\providecommand \bibfnamefont [1]{#1}%
\providecommand \citenamefont [1]{#1}%
\providecommand \href@noop [0]{\@secondoftwo}%
\providecommand \href [0]{\begingroup \@sanitize@url \@href}%
\providecommand \@href[1]{\@@startlink{#1}\@@href}%
\providecommand \@@href[1]{\endgroup#1\@@endlink}%
\providecommand \@sanitize@url [0]{\catcode `\\12\catcode `\$12\catcode
  `\&12\catcode `\#12\catcode `\^12\catcode `\_12\catcode `\%12\relax}%
\providecommand \@@startlink[1]{}%
\providecommand \@@endlink[0]{}%
\providecommand \url  [0]{\begingroup\@sanitize@url \@url }%
\providecommand \@url [1]{\endgroup\@href {#1}{\urlprefix }}%
\providecommand \urlprefix  [0]{URL }%
\providecommand \Eprint [0]{\href }%
\providecommand \doibase [0]{http://dx.doi.org/}%
\providecommand \selectlanguage [0]{\@gobble}%
\providecommand \bibinfo  [0]{\@secondoftwo}%
\providecommand \bibfield  [0]{\@secondoftwo}%
\providecommand \translation [1]{[#1]}%
\providecommand \BibitemOpen [0]{}%
\providecommand \bibitemStop [0]{}%
\providecommand \bibitemNoStop [0]{.\EOS\space}%
\providecommand \EOS [0]{\spacefactor3000\relax}%
\providecommand \BibitemShut  [1]{\csname bibitem#1\endcsname}%
\let\auto@bib@innerbib\@empty
\bibitem [{\citenamefont {Arsene}\ \emph {et~al.}(2005)\citenamefont {Arsene}
  \emph {et~al.}}]{Arsene:2004fa}%
  \BibitemOpen
  \bibfield  {author} {\bibinfo {author} {\bibfnamefont {I.}~\bibnamefont
  {Arsene}} \emph {et~al.} (\bibinfo {collaboration} {BRAHMS Collaboration}),\
  }\href {\doibase 10.1016/j.nuclphysa.2005.02.130} {\bibfield  {journal}
  {\bibinfo  {journal} {Nucl. Phys. A}\ }\textbf {\bibinfo {volume} {757}},\
  \bibinfo {pages} {1} (\bibinfo {year} {2005})},\ \Eprint
  {http://arxiv.org/abs/nucl-ex/0410020} {arXiv:nucl-ex/0410020} \BibitemShut
  {NoStop}%
\bibitem [{\citenamefont {Adcox}\ \emph {et~al.}(2005)\citenamefont {Adcox}
  \emph {et~al.}}]{Adcox:2004mh}%
  \BibitemOpen
  \bibfield  {author} {\bibinfo {author} {\bibfnamefont {K.}~\bibnamefont
  {Adcox}} \emph {et~al.} (\bibinfo {collaboration} {PHENIX Collaboration}),\
  }\href {\doibase 10.1016/j.nuclphysa.2005.03.086} {\bibfield  {journal}
  {\bibinfo  {journal} {Nucl. Phys. A}\ }\textbf {\bibinfo {volume} {757}},\
  \bibinfo {pages} {184} (\bibinfo {year} {2005})},\ \Eprint
  {http://arxiv.org/abs/nucl-ex/0410003} {arXiv:nucl-ex/0410003} \BibitemShut
  {NoStop}%
\bibitem [{\citenamefont {Back}\ \emph {et~al.}(2005)\citenamefont {Back} \emph
  {et~al.}}]{Back:2004je}%
  \BibitemOpen
  \bibfield  {author} {\bibinfo {author} {\bibfnamefont {B.~B.}\ \bibnamefont
  {Back}} \emph {et~al.} (\bibinfo {collaboration} {PHOBOS Collaboration}),\
  }\href {\doibase 10.1016/j.nuclphysa.2005.03.084} {\bibfield  {journal}
  {\bibinfo  {journal} {Nucl. Phys. A}\ }\textbf {\bibinfo {volume} {757}},\
  \bibinfo {pages} {28} (\bibinfo {year} {2005})},\ \Eprint
  {http://arxiv.org/abs/nucl-ex/0410022} {arXiv:nucl-ex/0410022} \BibitemShut
  {NoStop}%
\bibitem [{\citenamefont {Adams}\ \emph
  {et~al.}(2005{\natexlab{a}})\citenamefont {Adams} \emph
  {et~al.}}]{Adams:2005dq}%
  \BibitemOpen
  \bibfield  {author} {\bibinfo {author} {\bibfnamefont {J.}~\bibnamefont
  {Adams}} \emph {et~al.} (\bibinfo {collaboration} {STAR Collaboration}),\
  }\href {\doibase 10.1016/j.nuclphysa.2005.03.085} {\bibfield  {journal}
  {\bibinfo  {journal} {Nucl. Phys. A}\ }\textbf {\bibinfo {volume} {757}},\
  \bibinfo {pages} {102} (\bibinfo {year} {2005}{\natexlab{a}})},\ \Eprint
  {http://arxiv.org/abs/nucl-ex/0501009} {arXiv:nucl-ex/0501009} \BibitemShut
  {NoStop}%
\bibitem [{\citenamefont {Roland}\ \emph {et~al.}(2014)\citenamefont {Roland},
  \citenamefont {Safarik},\ and\ \citenamefont {Steinberg}}]{ROLAND201470}%
  \BibitemOpen
  \bibfield  {author} {\bibinfo {author} {\bibfnamefont {G.}~\bibnamefont
  {Roland}}, \bibinfo {author} {\bibfnamefont {K.}~\bibnamefont {Safarik}}, \
  and\ \bibinfo {author} {\bibfnamefont {P.}~\bibnamefont {Steinberg}},\ }\href
  {\doibase 10.1016/j.ppnp.2014.05.001} {\bibfield  {journal} {\bibinfo
  {journal} {Prog. Part. Nucl. Phys.}\ }\textbf {\bibinfo {volume} {77}},\
  \bibinfo {pages} {70} (\bibinfo {year} {2014})}\BibitemShut {NoStop}%
\bibitem [{\citenamefont {Voloshin}\ and\ \citenamefont
  {Zhang}(1996)}]{Voloshin:1994mz}%
  \BibitemOpen
  \bibfield  {author} {\bibinfo {author} {\bibfnamefont {S.}~\bibnamefont
  {Voloshin}}\ and\ \bibinfo {author} {\bibfnamefont {Y.}~\bibnamefont
  {Zhang}},\ }\href {\doibase 10.1007/s002880050141} {\bibfield  {journal}
  {\bibinfo  {journal} {Z. Phys. C}\ }\textbf {\bibinfo {volume} {70}},\
  \bibinfo {pages} {665} (\bibinfo {year} {1996})},\ \Eprint
  {http://arxiv.org/abs/hep-ph/9407282} {arXiv:hep-ph/9407282} \BibitemShut
  {NoStop}%
\bibitem [{\citenamefont {Poskanzer}\ and\ \citenamefont
  {Voloshin}(1998)}]{Poskanzer:1998yz}%
  \BibitemOpen
  \bibfield  {author} {\bibinfo {author} {\bibfnamefont {A.~M.}\ \bibnamefont
  {Poskanzer}}\ and\ \bibinfo {author} {\bibfnamefont {S.~A.}\ \bibnamefont
  {Voloshin}},\ }\href {\doibase 10.1103/PhysRevC.58.1671} {\bibfield
  {journal} {\bibinfo  {journal} {Phys. Rev. C}\ }\textbf {\bibinfo {volume}
  {58}},\ \bibinfo {pages} {1671} (\bibinfo {year} {1998})},\ \Eprint
  {http://arxiv.org/abs/nucl-ex/9805001} {arXiv:nucl-ex/9805001} \BibitemShut
  {NoStop}%
\bibitem [{\citenamefont {Qiu}\ and\ \citenamefont {Heinz}(2011)}]{Qiu:2011iv}%
  \BibitemOpen
  \bibfield  {author} {\bibinfo {author} {\bibfnamefont {Z.}~\bibnamefont
  {Qiu}}\ and\ \bibinfo {author} {\bibfnamefont {U.~W.}\ \bibnamefont
  {Heinz}},\ }\href {\doibase 10.1103/PhysRevC.84.024911} {\bibfield  {journal}
  {\bibinfo  {journal} {Phys. Rev. C}\ }\textbf {\bibinfo {volume} {84}},\
  \bibinfo {pages} {024911} (\bibinfo {year} {2011})},\ \Eprint
  {http://arxiv.org/abs/1104.0650} {arXiv:1104.0650 [nucl-th]} \BibitemShut
  {NoStop}%
\bibitem [{\citenamefont {Lacey}(2006)}]{Lacey:2005qq}%
  \BibitemOpen
  \bibfield  {author} {\bibinfo {author} {\bibfnamefont {R.~A.}\ \bibnamefont
  {Lacey}},\ }\href {\doibase 10.1016/j.nuclphysa.2006.06.041} {\bibfield
  {journal} {\bibinfo  {journal} {Nucl. Phys. A}\ }\textbf {\bibinfo {volume}
  {774}},\ \bibinfo {pages} {199} (\bibinfo {year} {2006})},\ \Eprint
  {http://arxiv.org/abs/nucl-ex/0510029} {arXiv:nucl-ex/0510029} \BibitemShut
  {NoStop}%
\bibitem [{\citenamefont {Song}\ \emph {et~al.}(2011)\citenamefont {Song},
  \citenamefont {Bass}, \citenamefont {Heinz}, \citenamefont {Hirano},\ and\
  \citenamefont {Shen}}]{Song:2010mg}%
  \BibitemOpen
  \bibfield  {author} {\bibinfo {author} {\bibfnamefont {H.}~\bibnamefont
  {Song}}, \bibinfo {author} {\bibfnamefont {S.~A.}\ \bibnamefont {Bass}},
  \bibinfo {author} {\bibfnamefont {U.}~\bibnamefont {Heinz}}, \bibinfo
  {author} {\bibfnamefont {T.}~\bibnamefont {Hirano}}, \ and\ \bibinfo {author}
  {\bibfnamefont {C.}~\bibnamefont {Shen}},\ }\href {\doibase
  10.1103/PhysRevLett.106.192301} {\bibfield  {journal} {\bibinfo  {journal}
  {Phys. Rev. Lett.}\ }\textbf {\bibinfo {volume} {106}},\ \bibinfo {pages}
  {192301} (\bibinfo {year} {2011})},\ \bibinfo {note} {[Erratum:
  Phys.Rev.Lett. 109, 139904 (2012)]},\ \Eprint
  {http://arxiv.org/abs/1011.2783} {arXiv:1011.2783 [nucl-th]} \BibitemShut
  {NoStop}%
\bibitem [{\citenamefont {Niemi}\ \emph {et~al.}(2013)\citenamefont {Niemi},
  \citenamefont {Denicol}, \citenamefont {Holopainen},\ and\ \citenamefont
  {Huovinen}}]{Niemi:2012aj}%
  \BibitemOpen
  \bibfield  {author} {\bibinfo {author} {\bibfnamefont {H.}~\bibnamefont
  {Niemi}}, \bibinfo {author} {\bibfnamefont {G.~S.}\ \bibnamefont {Denicol}},
  \bibinfo {author} {\bibfnamefont {H.}~\bibnamefont {Holopainen}}, \ and\
  \bibinfo {author} {\bibfnamefont {P.}~\bibnamefont {Huovinen}},\ }\href
  {\doibase 10.1103/PhysRevC.87.054901} {\bibfield  {journal} {\bibinfo
  {journal} {Phys. Rev. C}\ }\textbf {\bibinfo {volume} {87}},\ \bibinfo
  {pages} {054901} (\bibinfo {year} {2013})},\ \Eprint
  {http://arxiv.org/abs/1212.1008} {arXiv:1212.1008 [nucl-th]} \BibitemShut
  {NoStop}%
\bibitem [{\citenamefont {Gardim}\ \emph {et~al.}(2015)\citenamefont {Gardim},
  \citenamefont {Noronha-Hostler}, \citenamefont {Luzum},\ and\ \citenamefont
  {Grassi}}]{Gardim:2014tya}%
  \BibitemOpen
  \bibfield  {author} {\bibinfo {author} {\bibfnamefont {F.~G.}\ \bibnamefont
  {Gardim}}, \bibinfo {author} {\bibfnamefont {J.}~\bibnamefont
  {Noronha-Hostler}}, \bibinfo {author} {\bibfnamefont {M.}~\bibnamefont
  {Luzum}}, \ and\ \bibinfo {author} {\bibfnamefont {F.}~\bibnamefont
  {Grassi}},\ }\href {\doibase 10.1103/PhysRevC.91.034902} {\bibfield
  {journal} {\bibinfo  {journal} {Phys. Rev. C}\ }\textbf {\bibinfo {volume}
  {91}},\ \bibinfo {pages} {034902} (\bibinfo {year} {2015})},\ \Eprint
  {http://arxiv.org/abs/1411.2574} {arXiv:1411.2574 [nucl-th]} \BibitemShut
  {NoStop}%
\bibitem [{\citenamefont {Fu}(2015)}]{Fu:2015wba}%
  \BibitemOpen
  \bibfield  {author} {\bibinfo {author} {\bibfnamefont {J.}~\bibnamefont
  {Fu}},\ }\href {\doibase 10.1103/PhysRevC.92.024904} {\bibfield  {journal}
  {\bibinfo  {journal} {Phys. Rev. C}\ }\textbf {\bibinfo {volume} {92}},\
  \bibinfo {pages} {024904} (\bibinfo {year} {2015})}\BibitemShut {NoStop}%
\bibitem [{\citenamefont {Holopainen}\ \emph {et~al.}(2011)\citenamefont
  {Holopainen}, \citenamefont {Niemi},\ and\ \citenamefont
  {Eskola}}]{Holopainen:2010gz}%
  \BibitemOpen
  \bibfield  {author} {\bibinfo {author} {\bibfnamefont {H.}~\bibnamefont
  {Holopainen}}, \bibinfo {author} {\bibfnamefont {H.}~\bibnamefont {Niemi}}, \
  and\ \bibinfo {author} {\bibfnamefont {K.~J.}\ \bibnamefont {Eskola}},\
  }\href {\doibase 10.1103/PhysRevC.83.034901} {\bibfield  {journal} {\bibinfo
  {journal} {Phys. Rev. C}\ }\textbf {\bibinfo {volume} {83}},\ \bibinfo
  {pages} {034901} (\bibinfo {year} {2011})},\ \Eprint
  {http://arxiv.org/abs/1007.0368} {arXiv:1007.0368 [hep-ph]} \BibitemShut
  {NoStop}%
\bibitem [{\citenamefont {Qin}\ \emph {et~al.}(2010)\citenamefont {Qin},
  \citenamefont {Petersen}, \citenamefont {Bass},\ and\ \citenamefont
  {Muller}}]{Qin:2010pf}%
  \BibitemOpen
  \bibfield  {author} {\bibinfo {author} {\bibfnamefont {G.-Y.}\ \bibnamefont
  {Qin}}, \bibinfo {author} {\bibfnamefont {H.}~\bibnamefont {Petersen}},
  \bibinfo {author} {\bibfnamefont {S.~A.}\ \bibnamefont {Bass}}, \ and\
  \bibinfo {author} {\bibfnamefont {B.}~\bibnamefont {Muller}},\ }\href
  {\doibase 10.1103/PhysRevC.82.064903} {\bibfield  {journal} {\bibinfo
  {journal} {Phys. Rev. C}\ }\textbf {\bibinfo {volume} {82}},\ \bibinfo
  {pages} {064903} (\bibinfo {year} {2010})},\ \Eprint
  {http://arxiv.org/abs/1009.1847} {arXiv:1009.1847 [nucl-th]} \BibitemShut
  {NoStop}%
\bibitem [{\citenamefont {Gale}\ \emph {et~al.}(2013)\citenamefont {Gale},
  \citenamefont {Jeon}, \citenamefont {Schenke}, \citenamefont {Tribedy},\ and\
  \citenamefont {Venugopalan}}]{Gale:2012rq}%
  \BibitemOpen
  \bibfield  {author} {\bibinfo {author} {\bibfnamefont {C.}~\bibnamefont
  {Gale}}, \bibinfo {author} {\bibfnamefont {S.}~\bibnamefont {Jeon}}, \bibinfo
  {author} {\bibfnamefont {B.}~\bibnamefont {Schenke}}, \bibinfo {author}
  {\bibfnamefont {P.}~\bibnamefont {Tribedy}}, \ and\ \bibinfo {author}
  {\bibfnamefont {R.}~\bibnamefont {Venugopalan}},\ }\href {\doibase
  10.1103/PhysRevLett.110.012302} {\bibfield  {journal} {\bibinfo  {journal}
  {Phys. Rev. Lett.}\ }\textbf {\bibinfo {volume} {110}},\ \bibinfo {pages}
  {012302} (\bibinfo {year} {2013})},\ \Eprint {http://arxiv.org/abs/1209.6330}
  {arXiv:1209.6330 [nucl-th]} \BibitemShut {NoStop}%
\bibitem [{\citenamefont {Liu}\ and\ \citenamefont
  {Lacey}(2018{\natexlab{a}})}]{Liu:2018hjh}%
  \BibitemOpen
  \bibfield  {author} {\bibinfo {author} {\bibfnamefont {P.}~\bibnamefont
  {Liu}}\ and\ \bibinfo {author} {\bibfnamefont {R.~A.}\ \bibnamefont
  {Lacey}},\ }\href {\doibase 10.1103/PhysRevC.98.021902} {\bibfield  {journal}
  {\bibinfo  {journal} {Phys. Rev. C}\ }\textbf {\bibinfo {volume} {98}},\
  \bibinfo {pages} {021902} (\bibinfo {year} {2018}{\natexlab{a}})},\ \Eprint
  {http://arxiv.org/abs/1802.06595} {arXiv:1802.06595 [nucl-ex]} \BibitemShut
  {NoStop}%
\bibitem [{\citenamefont {Schenke}\ \emph {et~al.}(2011)\citenamefont
  {Schenke}, \citenamefont {Jeon},\ and\ \citenamefont
  {Gale}}]{Schenke:2011tv}%
  \BibitemOpen
  \bibfield  {author} {\bibinfo {author} {\bibfnamefont {B.}~\bibnamefont
  {Schenke}}, \bibinfo {author} {\bibfnamefont {S.}~\bibnamefont {Jeon}}, \
  and\ \bibinfo {author} {\bibfnamefont {C.}~\bibnamefont {Gale}},\ }\href
  {\doibase 10.1016/j.physletb.2011.06.065} {\bibfield  {journal} {\bibinfo
  {journal} {Phys. Lett. B}\ }\textbf {\bibinfo {volume} {702}},\ \bibinfo
  {pages} {59} (\bibinfo {year} {2011})},\ \Eprint
  {http://arxiv.org/abs/1102.0575} {arXiv:1102.0575 [hep-ph]} \BibitemShut
  {NoStop}%
\bibitem [{\citenamefont {Adare}\ \emph
  {et~al.}(2015{\natexlab{a}})\citenamefont {Adare} \emph
  {et~al.}}]{PHENIX:2014fnc}%
  \BibitemOpen
  \bibfield  {author} {\bibinfo {author} {\bibfnamefont {A.}~\bibnamefont
  {Adare}} \emph {et~al.} (\bibinfo {collaboration} {PHENIX Collaboration}),\
  }\href {\doibase 10.1103/PhysRevLett.114.192301} {\bibfield  {journal}
  {\bibinfo  {journal} {Phys. Rev. Lett.}\ }\textbf {\bibinfo {volume} {114}},\
  \bibinfo {pages} {192301} (\bibinfo {year} {2015}{\natexlab{a}})},\ \Eprint
  {http://arxiv.org/abs/1404.7461} {arXiv:1404.7461 [nucl-ex]} \BibitemShut
  {NoStop}%
\bibitem [{\citenamefont {Adare}\ \emph
  {et~al.}(2015{\natexlab{b}})\citenamefont {Adare} \emph
  {et~al.}}]{PHENIX:2015idk}%
  \BibitemOpen
  \bibfield  {author} {\bibinfo {author} {\bibfnamefont {A.}~\bibnamefont
  {Adare}} \emph {et~al.} (\bibinfo {collaboration} {PHENIX Collaboration}),\
  }\href {\doibase 10.1103/PhysRevLett.115.142301} {\bibfield  {journal}
  {\bibinfo  {journal} {Phys. Rev. Lett.}\ }\textbf {\bibinfo {volume} {115}},\
  \bibinfo {pages} {142301} (\bibinfo {year} {2015}{\natexlab{b}})},\ \Eprint
  {http://arxiv.org/abs/1507.06273} {arXiv:1507.06273 [nucl-ex]} \BibitemShut
  {NoStop}%
\bibitem [{\citenamefont {Aidala}\ \emph {et~al.}(2019)\citenamefont {Aidala}
  \emph {et~al.}}]{PHENIX:2018lia}%
  \BibitemOpen
  \bibfield  {author} {\bibinfo {author} {\bibfnamefont {C.}~\bibnamefont
  {Aidala}} \emph {et~al.} (\bibinfo {collaboration} {PHENIX Collaboration}),\
  }\href {\doibase 10.1038/s41567-018-0360-0} {\bibfield  {journal} {\bibinfo
  {journal} {Nature Phys.}\ }\textbf {\bibinfo {volume} {15}},\ \bibinfo
  {pages} {214} (\bibinfo {year} {2019})},\ \Eprint
  {http://arxiv.org/abs/1805.02973} {arXiv:1805.02973 [nucl-ex]} \BibitemShut
  {NoStop}%
\bibitem [{\citenamefont {Acharya}\ \emph
  {et~al.}(2022{\natexlab{a}})\citenamefont {Acharya} \emph
  {et~al.}}]{PHENIX:2021ubk}%
  \BibitemOpen
  \bibfield  {author} {\bibinfo {author} {\bibfnamefont {U.~A.}\ \bibnamefont
  {Acharya}} \emph {et~al.} (\bibinfo {collaboration} {PHENIX Collaboration}),\
  }\href {\doibase 10.1103/PhysRevC.105.024901} {\bibfield  {journal} {\bibinfo
   {journal} {Phys. Rev. C}\ }\textbf {\bibinfo {volume} {105}},\ \bibinfo
  {pages} {024901} (\bibinfo {year} {2022}{\natexlab{a}})},\ \Eprint
  {http://arxiv.org/abs/2107.06634} {arXiv:2107.06634 [hep-ex]} \BibitemShut
  {NoStop}%
\bibitem [{\citenamefont {Acharya}\ \emph
  {et~al.}(2022{\natexlab{b}})\citenamefont {Acharya} \emph
  {et~al.}}]{PHENIX:2022nht}%
  \BibitemOpen
  \bibfield  {author} {\bibinfo {author} {\bibfnamefont {U.~A.}\ \bibnamefont
  {Acharya}} \emph {et~al.} (\bibinfo {collaboration} {PHENIX Collaboration}),\
  }\href@noop {} {\  (\bibinfo {year} {2022}{\natexlab{b}})},\ \Eprint
  {http://arxiv.org/abs/2203.09894} {arXiv:2203.09894 [nucl-ex]} \BibitemShut
  {NoStop}%
\bibitem [{\citenamefont {Adamczyk}\ \emph
  {et~al.}(2015{\natexlab{a}})\citenamefont {Adamczyk} \emph
  {et~al.}}]{2015265}%
  \BibitemOpen
  \bibfield  {author} {\bibinfo {author} {\bibfnamefont {L.}~\bibnamefont
  {Adamczyk}} \emph {et~al.} (\bibinfo {collaboration} {STAR Collaboration}),\
  }\href {\doibase 10.1016/j.physletb.2015.05.075} {\bibfield  {journal}
  {\bibinfo  {journal} {Phys. Lett. B}\ }\textbf {\bibinfo {volume} {747}},\
  \bibinfo {pages} {265} (\bibinfo {year} {2015}{\natexlab{a}})},\ \Eprint
  {http://arxiv.org/abs/1502.07652} {arXiv:1502.07652 [nucl-ex]} \BibitemShut
  {NoStop}%
\bibitem [{\citenamefont {Adam}\ \emph {et~al.}(2019)\citenamefont {Adam} \emph
  {et~al.}}]{STAR:2019zaf}%
  \BibitemOpen
  \bibfield  {author} {\bibinfo {author} {\bibfnamefont {J.}~\bibnamefont
  {Adam}} \emph {et~al.} (\bibinfo {collaboration} {STAR Collaboration}),\
  }\href {\doibase 10.1103/PhysRevLett.122.172301} {\bibfield  {journal}
  {\bibinfo  {journal} {Phys. Rev. Lett.}\ }\textbf {\bibinfo {volume} {122}},\
  \bibinfo {pages} {172301} (\bibinfo {year} {2019})},\ \Eprint
  {http://arxiv.org/abs/1901.08155} {arXiv:1901.08155 [nucl-ex]} \BibitemShut
  {NoStop}%
\bibitem [{\citenamefont {Chatrchyan}\ \emph {et~al.}(2013)\citenamefont
  {Chatrchyan} \emph {et~al.}}]{Chatrchyan:2013nka}%
  \BibitemOpen
  \bibfield  {author} {\bibinfo {author} {\bibfnamefont {S.}~\bibnamefont
  {Chatrchyan}} \emph {et~al.} (\bibinfo {collaboration} {CMS Collaboration}),\
  }\href {\doibase 10.1016/j.physletb.2013.06.028} {\bibfield  {journal}
  {\bibinfo  {journal} {Phys. Lett. B}\ }\textbf {\bibinfo {volume} {724}},\
  \bibinfo {pages} {213} (\bibinfo {year} {2013})},\ \Eprint
  {http://arxiv.org/abs/1305.0609} {arXiv:1305.0609 [nucl-ex]} \BibitemShut
  {NoStop}%
\bibitem [{\citenamefont {Abelev}\ \emph {et~al.}(2013)\citenamefont {Abelev}
  \emph {et~al.}}]{Abelev:2012ola}%
  \BibitemOpen
  \bibfield  {author} {\bibinfo {author} {\bibfnamefont {B.}~\bibnamefont
  {Abelev}} \emph {et~al.} (\bibinfo {collaboration} {ALICE Collaboration}),\
  }\href {\doibase 10.1016/j.physletb.2013.01.012} {\bibfield  {journal}
  {\bibinfo  {journal} {Phys. Lett. B}\ }\textbf {\bibinfo {volume} {719}},\
  \bibinfo {pages} {29} (\bibinfo {year} {2013})},\ \Eprint
  {http://arxiv.org/abs/1212.2001} {arXiv:1212.2001 [nucl-ex]} \BibitemShut
  {NoStop}%
\bibitem [{\citenamefont {Aad}\ \emph {et~al.}(2013)\citenamefont {Aad} \emph
  {et~al.}}]{Aad:2012gla}%
  \BibitemOpen
  \bibfield  {author} {\bibinfo {author} {\bibfnamefont {G.}~\bibnamefont
  {Aad}} \emph {et~al.} (\bibinfo {collaboration} {ATLAS Collaboration}),\
  }\href {\doibase 10.1103/PhysRevLett.110.182302} {\bibfield  {journal}
  {\bibinfo  {journal} {Phys. Rev. Lett.}\ }\textbf {\bibinfo {volume} {110}},\
  \bibinfo {pages} {182302} (\bibinfo {year} {2013})},\ \Eprint
  {http://arxiv.org/abs/1212.5198} {arXiv:1212.5198 [hep-ex]} \BibitemShut
  {NoStop}%
\bibitem [{\citenamefont {Aaboud}\ \emph {et~al.}(2017)\citenamefont {Aaboud}
  \emph {et~al.}}]{Aaboud:2017acw}%
  \BibitemOpen
  \bibfield  {author} {\bibinfo {author} {\bibfnamefont {M.}~\bibnamefont
  {Aaboud}} \emph {et~al.} (\bibinfo {collaboration} {ATLAS Collaboration}),\
  }\href {\doibase 10.1140/epjc/s10052-017-4988-1} {\bibfield  {journal}
  {\bibinfo  {journal} {Eur. Phys. J. C}\ }\textbf {\bibinfo {volume} {77}},\
  \bibinfo {pages} {428} (\bibinfo {year} {2017})},\ \Eprint
  {http://arxiv.org/abs/1705.04176} {arXiv:1705.04176 [hep-ex]} \BibitemShut
  {NoStop}%
\bibitem [{\citenamefont {Chesler}(2016)}]{Chesler:2016ceu}%
  \BibitemOpen
  \bibfield  {author} {\bibinfo {author} {\bibfnamefont {P.~M.}\ \bibnamefont
  {Chesler}},\ }\href {\doibase 10.1007/JHEP03(2016)146} {\bibfield  {journal}
  {\bibinfo  {journal} {JHEP}\ }\textbf {\bibinfo {volume} {03}},\ \bibinfo
  {pages} {146} (\bibinfo {year} {2016})},\ \Eprint
  {http://arxiv.org/abs/1601.01583} {arXiv:1601.01583 [hep-th]} \BibitemShut
  {NoStop}%
\bibitem [{\citenamefont {Welsh}\ \emph {et~al.}(2016)\citenamefont {Welsh},
  \citenamefont {Singer},\ and\ \citenamefont {Heinz}}]{PhysRevC.94.024919}%
  \BibitemOpen
  \bibfield  {author} {\bibinfo {author} {\bibfnamefont {K.}~\bibnamefont
  {Welsh}}, \bibinfo {author} {\bibfnamefont {J.}~\bibnamefont {Singer}}, \
  and\ \bibinfo {author} {\bibfnamefont {U.~W.}\ \bibnamefont {Heinz}},\ }\href
  {\doibase 10.1103/PhysRevC.94.024919} {\bibfield  {journal} {\bibinfo
  {journal} {Phys. Rev. C}\ }\textbf {\bibinfo {volume} {94}},\ \bibinfo
  {pages} {024919} (\bibinfo {year} {2016})},\ \Eprint
  {http://arxiv.org/abs/1605.09418} {arXiv:1605.09418 [nucl-th]} \BibitemShut
  {NoStop}%
\bibitem [{\citenamefont {Schenke}\ \emph {et~al.}(2012)\citenamefont
  {Schenke}, \citenamefont {Tribedy},\ and\ \citenamefont
  {Venugopalan}}]{PhysRevLett.108.252301}%
  \BibitemOpen
  \bibfield  {author} {\bibinfo {author} {\bibfnamefont {B.}~\bibnamefont
  {Schenke}}, \bibinfo {author} {\bibfnamefont {P.}~\bibnamefont {Tribedy}}, \
  and\ \bibinfo {author} {\bibfnamefont {R.}~\bibnamefont {Venugopalan}},\
  }\href {\doibase 10.1103/PhysRevLett.108.252301} {\bibfield  {journal}
  {\bibinfo  {journal} {Phys. Rev. Lett.}\ }\textbf {\bibinfo {volume} {108}},\
  \bibinfo {pages} {252301} (\bibinfo {year} {2012})}\BibitemShut {NoStop}%
\bibitem [{\citenamefont {Nagle}\ \emph {et~al.}(2014)\citenamefont {Nagle},
  \citenamefont {Adare}, \citenamefont {Beckman}, \citenamefont {Koblesky},
  \citenamefont {Orjuela~Koop}, \citenamefont {McGlinchey}, \citenamefont
  {Romatschke}, \citenamefont {Carlson}, \citenamefont {Lynn},\ and\
  \citenamefont {McCumber}}]{PhysRevLett.113.112301}%
  \BibitemOpen
  \bibfield  {author} {\bibinfo {author} {\bibfnamefont {J.~L.}\ \bibnamefont
  {Nagle}}, \bibinfo {author} {\bibfnamefont {A.}~\bibnamefont {Adare}},
  \bibinfo {author} {\bibfnamefont {S.}~\bibnamefont {Beckman}}, \bibinfo
  {author} {\bibfnamefont {T.}~\bibnamefont {Koblesky}}, \bibinfo {author}
  {\bibfnamefont {J.}~\bibnamefont {Orjuela~Koop}}, \bibinfo {author}
  {\bibfnamefont {D.}~\bibnamefont {McGlinchey}}, \bibinfo {author}
  {\bibfnamefont {P.}~\bibnamefont {Romatschke}}, \bibinfo {author}
  {\bibfnamefont {J.}~\bibnamefont {Carlson}}, \bibinfo {author} {\bibfnamefont
  {J.~E.}\ \bibnamefont {Lynn}}, \ and\ \bibinfo {author} {\bibfnamefont
  {M.}~\bibnamefont {McCumber}},\ }\href {\doibase
  10.1103/PhysRevLett.113.112301} {\bibfield  {journal} {\bibinfo  {journal}
  {Phys. Rev. Lett.}\ }\textbf {\bibinfo {volume} {113}},\ \bibinfo {pages}
  {112301} (\bibinfo {year} {2014})},\ \Eprint {http://arxiv.org/abs/1312.4565}
  {arXiv:1312.4565 [nucl-th]} \BibitemShut {NoStop}%
\bibitem [{\citenamefont {Bzdak}\ and\ \citenamefont
  {Ma}(2014)}]{Bzdak:2014dia}%
  \BibitemOpen
  \bibfield  {author} {\bibinfo {author} {\bibfnamefont {A.}~\bibnamefont
  {Bzdak}}\ and\ \bibinfo {author} {\bibfnamefont {G.-L.}\ \bibnamefont {Ma}},\
  }\href {\doibase 10.1103/PhysRevLett.113.252301} {\bibfield  {journal}
  {\bibinfo  {journal} {Phys. Rev. Lett.}\ }\textbf {\bibinfo {volume} {113}},\
  \bibinfo {pages} {252301} (\bibinfo {year} {2014})},\ \Eprint
  {http://arxiv.org/abs/1406.2804} {arXiv:1406.2804 [hep-ph]} \BibitemShut
  {NoStop}%
\bibitem [{\citenamefont {Zhou}\ \emph {et~al.}(2015)\citenamefont {Zhou},
  \citenamefont {Zhu}, \citenamefont {Li},\ and\ \citenamefont
  {Song}}]{Zhou:2015iba}%
  \BibitemOpen
  \bibfield  {author} {\bibinfo {author} {\bibfnamefont {Y.}~\bibnamefont
  {Zhou}}, \bibinfo {author} {\bibfnamefont {X.}~\bibnamefont {Zhu}}, \bibinfo
  {author} {\bibfnamefont {P.}~\bibnamefont {Li}}, \ and\ \bibinfo {author}
  {\bibfnamefont {H.}~\bibnamefont {Song}},\ }\href {\doibase
  10.1103/PhysRevC.91.064908} {\bibfield  {journal} {\bibinfo  {journal} {Phys.
  Rev. C}\ }\textbf {\bibinfo {volume} {91}},\ \bibinfo {pages} {064908}
  (\bibinfo {year} {2015})},\ \Eprint {http://arxiv.org/abs/1503.06986}
  {arXiv:1503.06986 [nucl-th]} \BibitemShut {NoStop}%
\bibitem [{\citenamefont
  {Romatschke}(2015{\natexlab{a}})}]{Romatschke:2015dha}%
  \BibitemOpen
  \bibfield  {author} {\bibinfo {author} {\bibfnamefont {P.}~\bibnamefont
  {Romatschke}},\ }\href {\doibase 10.1140/epjc/s10052-015-3646-8} {\bibfield
  {journal} {\bibinfo  {journal} {Eur. Phys. J. C}\ }\textbf {\bibinfo {volume}
  {75}},\ \bibinfo {pages} {429} (\bibinfo {year} {2015}{\natexlab{a}})},\
  \Eprint {http://arxiv.org/abs/1504.02529} {arXiv:1504.02529 [nucl-th]}
  \BibitemShut {NoStop}%
\bibitem [{\citenamefont {Bierlich}\ \emph {et~al.}(2018)\citenamefont
  {Bierlich}, \citenamefont {Gustafson},\ and\ \citenamefont
  {L\"onnblad}}]{Bierlich:2017vhg}%
  \BibitemOpen
  \bibfield  {author} {\bibinfo {author} {\bibfnamefont {C.}~\bibnamefont
  {Bierlich}}, \bibinfo {author} {\bibfnamefont {G.}~\bibnamefont {Gustafson}},
  \ and\ \bibinfo {author} {\bibfnamefont {L.}~\bibnamefont {L\"onnblad}},\
  }\href {\doibase 10.1016/j.physletb.2018.01.069} {\bibfield  {journal}
  {\bibinfo  {journal} {Phys. Lett. B}\ }\textbf {\bibinfo {volume} {779}},\
  \bibinfo {pages} {58} (\bibinfo {year} {2018})},\ \Eprint
  {http://arxiv.org/abs/1710.09725} {arXiv:1710.09725 [hep-ph]} \BibitemShut
  {NoStop}%
\bibitem [{\citenamefont {Mace}\ \emph {et~al.}(2018)\citenamefont {Mace},
  \citenamefont {Skokov}, \citenamefont {Tribedy},\ and\ \citenamefont
  {Venugopalan}}]{Mark:2018}%
  \BibitemOpen
  \bibfield  {author} {\bibinfo {author} {\bibfnamefont {M.}~\bibnamefont
  {Mace}}, \bibinfo {author} {\bibfnamefont {V.~V.}\ \bibnamefont {Skokov}},
  \bibinfo {author} {\bibfnamefont {P.}~\bibnamefont {Tribedy}}, \ and\
  \bibinfo {author} {\bibfnamefont {R.}~\bibnamefont {Venugopalan}},\ }\href
  {\doibase 10.1103/PhysRevLett.121.052301} {\bibfield  {journal} {\bibinfo
  {journal} {Phys. Rev. Lett.}\ }\textbf {\bibinfo {volume} {121}},\ \bibinfo
  {pages} {052301} (\bibinfo {year} {2018})},\ \bibinfo {note} {[Erratum:
  Phys.Rev.Lett. 123, 039901 (2019)]},\ \Eprint
  {http://arxiv.org/abs/1805.09342} {arXiv:1805.09342 [hep-ph]} \BibitemShut
  {NoStop}%
\bibitem [{\citenamefont {Dusling}\ and\ \citenamefont
  {Venugopalan}(2013)}]{Dusling:2013oia}%
  \BibitemOpen
  \bibfield  {author} {\bibinfo {author} {\bibfnamefont {K.}~\bibnamefont
  {Dusling}}\ and\ \bibinfo {author} {\bibfnamefont {R.}~\bibnamefont
  {Venugopalan}},\ }\href {\doibase 10.1103/PhysRevD.87.094034} {\bibfield
  {journal} {\bibinfo  {journal} {Phys. Rev. D}\ }\textbf {\bibinfo {volume}
  {87}},\ \bibinfo {pages} {094034} (\bibinfo {year} {2013})},\ \Eprint
  {http://arxiv.org/abs/1302.7018} {arXiv:1302.7018 [hep-ph]} \BibitemShut
  {NoStop}%
\bibitem [{\citenamefont {Nie}\ \emph {et~al.}(2019)\citenamefont {Nie},
  \citenamefont {Yi}, \citenamefont {Ma},\ and\ \citenamefont
  {Jia}}]{PhysRevC.100.064905}%
  \BibitemOpen
  \bibfield  {author} {\bibinfo {author} {\bibfnamefont {M.}~\bibnamefont
  {Nie}}, \bibinfo {author} {\bibfnamefont {L.}~\bibnamefont {Yi}}, \bibinfo
  {author} {\bibfnamefont {G.}~\bibnamefont {Ma}}, \ and\ \bibinfo {author}
  {\bibfnamefont {J.}~\bibnamefont {Jia}},\ }\href {\doibase
  10.1103/PhysRevC.100.064905} {\bibfield  {journal} {\bibinfo  {journal}
  {Phys. Rev. C}\ }\textbf {\bibinfo {volume} {100}},\ \bibinfo {pages}
  {064905} (\bibinfo {year} {2019})},\ \Eprint
  {http://arxiv.org/abs/1906.01422} {arXiv:1906.01422 [nucl-th]} \BibitemShut
  {NoStop}%
\bibitem [{\citenamefont {Denicol}\ \emph {et~al.}(2012)\citenamefont
  {Denicol}, \citenamefont {Niemi}, \citenamefont {Molnar},\ and\ \citenamefont
  {Rischke}}]{Denicol:2012cn}%
  \BibitemOpen
  \bibfield  {author} {\bibinfo {author} {\bibfnamefont {G.~S.}\ \bibnamefont
  {Denicol}}, \bibinfo {author} {\bibfnamefont {H.}~\bibnamefont {Niemi}},
  \bibinfo {author} {\bibfnamefont {E.}~\bibnamefont {Molnar}}, \ and\ \bibinfo
  {author} {\bibfnamefont {D.~H.}\ \bibnamefont {Rischke}},\ }\href {\doibase
  10.1103/PhysRevD.85.114047} {\bibfield  {journal} {\bibinfo  {journal} {Phys.
  Rev. D}\ }\textbf {\bibinfo {volume} {85}},\ \bibinfo {pages} {114047}
  (\bibinfo {year} {2012})},\ \bibinfo {note} {[Erratum: Phys.Rev.D 91, 039902
  (2015)]},\ \Eprint {http://arxiv.org/abs/1202.4551} {arXiv:1202.4551
  [nucl-th]} \BibitemShut {NoStop}%
\bibitem [{\citenamefont {Florkowski}\ \emph {et~al.}(2016)\citenamefont
  {Florkowski}, \citenamefont {Ryblewski},\ and\ \citenamefont
  {Spali\'nski}}]{Florkowski:2016zsi}%
  \BibitemOpen
  \bibfield  {author} {\bibinfo {author} {\bibfnamefont {W.}~\bibnamefont
  {Florkowski}}, \bibinfo {author} {\bibfnamefont {R.}~\bibnamefont
  {Ryblewski}}, \ and\ \bibinfo {author} {\bibfnamefont {M.}~\bibnamefont
  {Spali\'nski}},\ }\href {\doibase 10.1103/PhysRevD.94.114025} {\bibfield
  {journal} {\bibinfo  {journal} {Phys. Rev. D}\ }\textbf {\bibinfo {volume}
  {94}},\ \bibinfo {pages} {114025} (\bibinfo {year} {2016})},\ \Eprint
  {http://arxiv.org/abs/1608.07558} {arXiv:1608.07558 [nucl-th]} \BibitemShut
  {NoStop}%
\bibitem [{\citenamefont {Romatschke}(2015{\natexlab{b}})}]{sonic}%
  \BibitemOpen
  \bibfield  {author} {\bibinfo {author} {\bibfnamefont {P.}~\bibnamefont
  {Romatschke}},\ }\href {\doibase 10.1140/epjc/s10052-015-3509-3} {\bibfield
  {journal} {\bibinfo  {journal} {Eur. Phys. J. C}\ }\textbf {\bibinfo {volume}
  {75}},\ \bibinfo {pages} {305} (\bibinfo {year} {2015}{\natexlab{b}})},\
  \Eprint {http://arxiv.org/abs/1502.04745} {arXiv:1502.04745 [nucl-th]}
  \BibitemShut {NoStop}%
\bibitem [{\citenamefont {Adams}\ \emph {et~al.}(2004)\citenamefont {Adams}
  \emph {et~al.}}]{adams:2004ja}%
  \BibitemOpen
  \bibfield  {author} {\bibinfo {author} {\bibfnamefont {J.}~\bibnamefont
  {Adams}} \emph {et~al.} (\bibinfo {collaboration} {STAR Collaboration}),\
  }\href {\doibase 10.1103/PhysRevLett.93.252301} {\bibfield  {journal}
  {\bibinfo  {journal} {Phys. Rev. Lett.}\ }\textbf {\bibinfo {volume} {93}},\
  \bibinfo {pages} {252301} (\bibinfo {year} {2004})},\ \Eprint
  {http://arxiv.org/abs/nucl-ex/0407007} {arXiv:nucl-ex/0407007} \BibitemShut
  {NoStop}%
\bibitem [{\citenamefont {Adams}\ \emph
  {et~al.}(2005{\natexlab{b}})\citenamefont {Adams} \emph
  {et~al.}}]{Adams2005:aj}%
  \BibitemOpen
  \bibfield  {author} {\bibinfo {author} {\bibfnamefont {J.}~\bibnamefont
  {Adams}} \emph {et~al.} (\bibinfo {collaboration} {STAR Collaboration}),\
  }\href {\doibase 10.1103/PhysRevC.72.014904} {\bibfield  {journal} {\bibinfo
  {journal} {Phys. Rev. C}\ }\textbf {\bibinfo {volume} {72}},\ \bibinfo
  {pages} {014904} (\bibinfo {year} {2005}{\natexlab{b}})},\ \Eprint
  {http://arxiv.org/abs/nucl-ex/0409033} {arXiv:nucl-ex/0409033} \BibitemShut
  {NoStop}%
\bibitem [{\citenamefont {Khachatryan}\ \emph {et~al.}(2017)\citenamefont
  {Khachatryan} \emph {et~al.}}]{2017193}%
  \BibitemOpen
  \bibfield  {author} {\bibinfo {author} {\bibfnamefont {V.}~\bibnamefont
  {Khachatryan}} \emph {et~al.} (\bibinfo {collaboration} {CMS
  Collaboration}),\ }\href {\doibase 10.1016/j.physletb.2016.12.009} {\bibfield
   {journal} {\bibinfo  {journal} {Phys. Lett. B}\ }\textbf {\bibinfo {volume}
  {765}},\ \bibinfo {pages} {193} (\bibinfo {year} {2017})},\ \Eprint
  {http://arxiv.org/abs/1606.06198} {arXiv:1606.06198 [nucl-ex]} \BibitemShut
  {NoStop}%
\bibitem [{\citenamefont {Adamczyk}\ \emph
  {et~al.}(2015{\natexlab{b}})\citenamefont {Adamczyk} \emph
  {et~al.}}]{2015333}%
  \BibitemOpen
  \bibfield  {author} {\bibinfo {author} {\bibfnamefont {L.}~\bibnamefont
  {Adamczyk}} \emph {et~al.} (\bibinfo {collaboration} {STAR Collaboration}),\
  }\href {\doibase 10.1016/j.physletb.2015.02.068} {\bibfield  {journal}
  {\bibinfo  {journal} {Phys. Lett. B}\ }\textbf {\bibinfo {volume} {743}},\
  \bibinfo {pages} {333} (\bibinfo {year} {2015}{\natexlab{b}})},\ \Eprint
  {http://arxiv.org/abs/1412.8437} {arXiv:1412.8437 [nucl-ex]} \BibitemShut
  {NoStop}%
\bibitem [{\citenamefont {Aad}\ \emph {et~al.}(2014)\citenamefont {Aad} \emph
  {et~al.}}]{PhysRevC.90.044906}%
  \BibitemOpen
  \bibfield  {author} {\bibinfo {author} {\bibfnamefont {G.}~\bibnamefont
  {Aad}} \emph {et~al.} (\bibinfo {collaboration} {ATLAS Collaboration}),\
  }\href {\doibase 10.1103/PhysRevC.90.044906} {\bibfield  {journal} {\bibinfo
  {journal} {Phys. Rev. C}\ }\textbf {\bibinfo {volume} {90}},\ \bibinfo
  {pages} {044906} (\bibinfo {year} {2014})},\ \Eprint
  {http://arxiv.org/abs/1409.1792} {arXiv:1409.1792 [hep-ex]} \BibitemShut
  {NoStop}%
\bibitem [{\citenamefont {Adare}\ \emph {et~al.}(2018)\citenamefont {Adare}
  \emph {et~al.}}]{PhysRevC.98.014912}%
  \BibitemOpen
  \bibfield  {author} {\bibinfo {author} {\bibfnamefont {A.}~\bibnamefont
  {Adare}} \emph {et~al.} (\bibinfo {collaboration} {PHENIX Collaboration}),\
  }\href {\doibase 10.1103/PhysRevC.98.014912} {\bibfield  {journal} {\bibinfo
  {journal} {Phys. Rev. C}\ }\textbf {\bibinfo {volume} {98}},\ \bibinfo
  {pages} {014912} (\bibinfo {year} {2018})},\ \Eprint
  {http://arxiv.org/abs/1711.09003} {arXiv:1711.09003 [hep-ex]} \BibitemShut
  {NoStop}%
\bibitem [{\citenamefont {Aad}\ \emph {et~al.}(2016)\citenamefont {Aad} \emph
  {et~al.}}]{Aad:2016}%
  \BibitemOpen
  \bibfield  {author} {\bibinfo {author} {\bibfnamefont {G.}~\bibnamefont
  {Aad}} \emph {et~al.} (\bibinfo {collaboration} {ATLAS Collaboration}),\
  }\href {\doibase 10.1103/PhysRevLett.116.172301} {\bibfield  {journal}
  {\bibinfo  {journal} {Phys. Rev. Lett.}\ }\textbf {\bibinfo {volume} {116}},\
  \bibinfo {pages} {172301} (\bibinfo {year} {2016})},\ \Eprint
  {http://arxiv.org/abs/1509.04776} {arXiv:1509.04776 [hep-ex]} \BibitemShut
  {NoStop}%
\bibitem [{\citenamefont {Llope}\ \emph {et~al.}(2014)\citenamefont {Llope}
  \emph {et~al.}}]{STARVPD}%
  \BibitemOpen
  \bibfield  {author} {\bibinfo {author} {\bibfnamefont {W.~J.}\ \bibnamefont
  {Llope}} \emph {et~al.},\ }\href {\doibase 10.1016/j.nima.2014.04.080}
  {\bibfield  {journal} {\bibinfo  {journal} {Nucl. Instrum. Meth.}\ }\textbf
  {\bibinfo {volume} {A759}},\ \bibinfo {pages} {23} (\bibinfo {year}
  {2014})}\BibitemShut {NoStop}%
\bibitem [{\citenamefont {Bieser}\ \emph {et~al.}(2003)\citenamefont {Bieser}
  \emph {et~al.}}]{STARtrig}%
  \BibitemOpen
  \bibfield  {author} {\bibinfo {author} {\bibfnamefont {F.~S.}\ \bibnamefont
  {Bieser}} \emph {et~al.},\ }\href {\doibase 10.1016/S0168-9002(02)01974-5}
  {\bibfield  {journal} {\bibinfo  {journal} {Nucl. Instrum. Meth.}\ }\textbf
  {\bibinfo {volume} {A499}},\ \bibinfo {pages} {766} (\bibinfo {year}
  {2003})}\BibitemShut {NoStop}%
\bibitem [{\citenamefont {Adler}\ \emph {et~al.}(2003)\citenamefont {Adler},
  \citenamefont {Denisov}, \citenamefont {Garcia}, \citenamefont {Murray},
  \citenamefont {Strobele},\ and\ \citenamefont {White}}]{STARZDC}%
  \BibitemOpen
  \bibfield  {author} {\bibinfo {author} {\bibfnamefont {C.}~\bibnamefont
  {Adler}}, \bibinfo {author} {\bibfnamefont {A.}~\bibnamefont {Denisov}},
  \bibinfo {author} {\bibfnamefont {E.}~\bibnamefont {Garcia}}, \bibinfo
  {author} {\bibfnamefont {M.}~\bibnamefont {Murray}}, \bibinfo {author}
  {\bibfnamefont {H.}~\bibnamefont {Strobele}}, \ and\ \bibinfo {author}
  {\bibfnamefont {S.}~\bibnamefont {White}},\ }\href {\doibase
  10.1016/j.nima.2003.08.112} {\bibfield  {journal} {\bibinfo  {journal} {Nucl.
  Instrum. Meth. A}\ }\textbf {\bibinfo {volume} {499}},\ \bibinfo {pages}
  {433} (\bibinfo {year} {2003})}\BibitemShut {NoStop}%
\bibitem [{\citenamefont {Llope}(2012)}]{STARTOF}%
  \BibitemOpen
  \bibfield  {author} {\bibinfo {author} {\bibfnamefont {W.~J.}\ \bibnamefont
  {Llope}} (\bibinfo {collaboration} {STAR}),\ }\href {\doibase
  10.1016/j.nima.2010.07.086} {\bibfield  {journal} {\bibinfo  {journal} {Nucl.
  Instrum. Meth. A}\ }\textbf {\bibinfo {volume} {661}},\ \bibinfo {pages}
  {S110} (\bibinfo {year} {2012})}\BibitemShut {NoStop}%
\bibitem [{\citenamefont {Adamczyk}\ \emph {et~al.}(2012)\citenamefont
  {Adamczyk} \emph {et~al.}}]{STARGlauber}%
  \BibitemOpen
  \bibfield  {author} {\bibinfo {author} {\bibfnamefont {L.}~\bibnamefont
  {Adamczyk}} \emph {et~al.} (\bibinfo {collaboration} {STAR Collaboration}),\
  }\href {\doibase 10.1103/PhysRevC.86.054908} {\bibfield  {journal} {\bibinfo
  {journal} {Phys. Rev. C}\ }\textbf {\bibinfo {volume} {86}},\ \bibinfo
  {pages} {054908} (\bibinfo {year} {2012})},\ \Eprint
  {http://arxiv.org/abs/1206.5528} {arXiv:1206.5528 [nucl-ex]} \BibitemShut
  {NoStop}%
\bibitem [{\citenamefont {Loizides}\ \emph {et~al.}(2015)\citenamefont
  {Loizides}, \citenamefont {Nagle},\ and\ \citenamefont
  {Steinberg}}]{Loizides_2015}%
  \BibitemOpen
  \bibfield  {author} {\bibinfo {author} {\bibfnamefont {C.}~\bibnamefont
  {Loizides}}, \bibinfo {author} {\bibfnamefont {J.}~\bibnamefont {Nagle}}, \
  and\ \bibinfo {author} {\bibfnamefont {P.}~\bibnamefont {Steinberg}},\ }\href
  {\doibase 10.1016/j.softx.2015.05.001} {\bibfield  {journal} {\bibinfo
  {journal} {{SoftwareX}}\ }\textbf {\bibinfo {volume} {1-2}},\ \bibinfo
  {pages} {13} (\bibinfo {year} {2015})}\BibitemShut {NoStop}%
\bibitem [{\citenamefont {Agostinelli}\ \emph {et~al.}(2003)\citenamefont
  {Agostinelli} \emph {et~al.}}]{GEANT}%
  \BibitemOpen
  \bibfield  {author} {\bibinfo {author} {\bibfnamefont {S.}~\bibnamefont
  {Agostinelli}} \emph {et~al.},\ }\href {\doibase
  https://doi.org/10.1016/S0168-9002(03)01368-8} {\bibfield  {journal}
  {\bibinfo  {journal} {Nucl. Instrum. Meth. A}\ }\textbf {\bibinfo {volume}
  {506}},\ \bibinfo {pages} {250 } (\bibinfo {year} {2003})}\BibitemShut
  {NoStop}%
\bibitem [{\citenamefont {Fine}\ \emph {et~al.}(2001)\citenamefont {Fine},
  \citenamefont {Fisyak}, \citenamefont {Perevoztchikov},\ and\ \citenamefont
  {Wenaus}}]{FINE200176}%
  \BibitemOpen
  \bibfield  {author} {\bibinfo {author} {\bibfnamefont {V.}~\bibnamefont
  {Fine}}, \bibinfo {author} {\bibfnamefont {Y.}~\bibnamefont {Fisyak}},
  \bibinfo {author} {\bibfnamefont {V.}~\bibnamefont {Perevoztchikov}}, \ and\
  \bibinfo {author} {\bibfnamefont {T.}~\bibnamefont {Wenaus}},\ }\href
  {\doibase https://doi.org/10.1016/S0010-4655(01)00257-0} {\bibfield
  {journal} {\bibinfo  {journal} {Computer Physics Communications}\ }\textbf
  {\bibinfo {volume} {140}},\ \bibinfo {pages} {76} (\bibinfo {year} {2001})},\
  \bibinfo {note} {cHEP2000}\BibitemShut {NoStop}%
\bibitem [{\citenamefont {Anderson}\ \emph {et~al.}(2003)\citenamefont
  {Anderson} \emph {et~al.}}]{STARTPC}%
  \BibitemOpen
  \bibfield  {author} {\bibinfo {author} {\bibfnamefont {M.}~\bibnamefont
  {Anderson}} \emph {et~al.},\ }\href {\doibase 10.1016/S0168-9002(02)01964-2}
  {\bibfield  {journal} {\bibinfo  {journal} {Nucl. Instrum. Meth. A}\ }\textbf
  {\bibinfo {volume} {499}},\ \bibinfo {pages} {659} (\bibinfo {year}
  {2003})},\ \Eprint {http://arxiv.org/abs/nucl-ex/0301015}
  {arXiv:nucl-ex/0301015} \BibitemShut {NoStop}%
\bibitem [{\citenamefont {Szelezniak}(2015)}]{STARHFT}%
  \BibitemOpen
  \bibfield  {author} {\bibinfo {author} {\bibfnamefont {M.}~\bibnamefont
  {Szelezniak}} (\bibinfo {collaboration} {STAR Collaboration}),\ }\href
  {\doibase 10.22323/1.227.0015} {\bibfield  {journal} {\bibinfo  {journal}
  {PoS}\ }\textbf {\bibinfo {volume} {Vertex2014}},\ \bibinfo {pages} {015}
  (\bibinfo {year} {2015})}\BibitemShut {NoStop}%
\bibitem [{\citenamefont {Borghini}\ \emph {et~al.}(2001)\citenamefont
  {Borghini}, \citenamefont {Dinh},\ and\ \citenamefont
  {Ollitrault}}]{PhysRevC.64.054901}%
  \BibitemOpen
  \bibfield  {author} {\bibinfo {author} {\bibfnamefont {N.}~\bibnamefont
  {Borghini}}, \bibinfo {author} {\bibfnamefont {P.~M.}\ \bibnamefont {Dinh}},
  \ and\ \bibinfo {author} {\bibfnamefont {J.-Y.}\ \bibnamefont {Ollitrault}},\
  }\href {\doibase 10.1103/PhysRevC.64.054901} {\bibfield  {journal} {\bibinfo
  {journal} {Phys. Rev. C}\ }\textbf {\bibinfo {volume} {64}},\ \bibinfo
  {pages} {054901} (\bibinfo {year} {2001})}\BibitemShut {NoStop}%
\bibitem [{\citenamefont {Gyulassy}\ and\ \citenamefont {Wang}(1994)}]{HIJING}%
  \BibitemOpen
  \bibfield  {author} {\bibinfo {author} {\bibfnamefont {M.}~\bibnamefont
  {Gyulassy}}\ and\ \bibinfo {author} {\bibfnamefont {X.-N.}\ \bibnamefont
  {Wang}},\ }\href {\doibase 10.1016/0010-4655(94)90057-4} {\bibfield
  {journal} {\bibinfo  {journal} {Comput. Phys. Commun.}\ }\textbf {\bibinfo
  {volume} {83}},\ \bibinfo {pages} {307} (\bibinfo {year} {1994})},\ \Eprint
  {http://arxiv.org/abs/nucl-th/9502021} {arXiv:nucl-th/9502021} \BibitemShut
  {NoStop}%
\bibitem [{\citenamefont {Zhao}\ \emph {et~al.}(2022)\citenamefont {Zhao},
  \citenamefont {Ryu}, \citenamefont {Shen},\ and\ \citenamefont
  {Schenke}}]{Zhao:2022ugy}%
  \BibitemOpen
  \bibfield  {author} {\bibinfo {author} {\bibfnamefont {W.}~\bibnamefont
  {Zhao}}, \bibinfo {author} {\bibfnamefont {S.}~\bibnamefont {Ryu}}, \bibinfo
  {author} {\bibfnamefont {C.}~\bibnamefont {Shen}}, \ and\ \bibinfo {author}
  {\bibfnamefont {B.}~\bibnamefont {Schenke}},\ }\href@noop {} {\  (\bibinfo
  {year} {2022})},\ \Eprint {http://arxiv.org/abs/2211.16376} {arXiv:2211.16376
  [nucl-th]} \BibitemShut {NoStop}%
\bibitem [{\citenamefont {Schenke}\ \emph
  {et~al.}(2020{\natexlab{a}})\citenamefont {Schenke}, \citenamefont {Shen},\
  and\ \citenamefont {Tribedy}}]{Schenke_2020}%
  \BibitemOpen
  \bibfield  {author} {\bibinfo {author} {\bibfnamefont {B.}~\bibnamefont
  {Schenke}}, \bibinfo {author} {\bibfnamefont {C.}~\bibnamefont {Shen}}, \
  and\ \bibinfo {author} {\bibfnamefont {P.}~\bibnamefont {Tribedy}},\ }\href
  {\doibase 10.1016/j.physletb.2020.135322} {\bibfield  {journal} {\bibinfo
  {journal} {Phys. Lett. B}\ }\textbf {\bibinfo {volume} {803}},\ \bibinfo
  {pages} {135322} (\bibinfo {year} {2020}{\natexlab{a}})},\ \Eprint
  {http://arxiv.org/abs/1908.06212} {arXiv:1908.06212 [nucl-th]} \BibitemShut
  {NoStop}%
\bibitem [{\citenamefont {Schenke}\ \emph
  {et~al.}(2020{\natexlab{b}})\citenamefont {Schenke}, \citenamefont {Shen},\
  and\ \citenamefont {Tribedy}}]{schenke2020running}%
  \BibitemOpen
  \bibfield  {author} {\bibinfo {author} {\bibfnamefont {B.}~\bibnamefont
  {Schenke}}, \bibinfo {author} {\bibfnamefont {C.}~\bibnamefont {Shen}}, \
  and\ \bibinfo {author} {\bibfnamefont {P.}~\bibnamefont {Tribedy}},\ }\href
  {\doibase 10.1103/PhysRevC.102.044905} {\bibfield  {journal} {\bibinfo
  {journal} {Phys. Rev. C}\ }\textbf {\bibinfo {volume} {102}},\ \bibinfo
  {pages} {044905} (\bibinfo {year} {2020}{\natexlab{b}})},\ \Eprint
  {http://arxiv.org/abs/2005.14682} {arXiv:2005.14682 [nucl-th]} \BibitemShut
  {NoStop}%
\bibitem [{\citenamefont {Liu}\ and\ \citenamefont
  {Lacey}(2018{\natexlab{b}})}]{Liu:2018xae}%
  \BibitemOpen
  \bibfield  {author} {\bibinfo {author} {\bibfnamefont {P.}~\bibnamefont
  {Liu}}\ and\ \bibinfo {author} {\bibfnamefont {R.~A.}\ \bibnamefont
  {Lacey}},\ }\href@noop {} {\  (\bibinfo {year} {2018}{\natexlab{b}})},\
  \Eprint {http://arxiv.org/abs/1804.04618} {arXiv:1804.04618 [nucl-ex]}
  \BibitemShut {NoStop}%
\end{thebibliography}%
\bibliographystyle{apsrev4-1}
\clearpage
\section{Supplement}

In this supplement, we present the STAR measurements of azimuthal anisotropy coefficients $v_{2,3}$ in the \heau and \dau collisions with centrality defined by the Au-going side Beam-Beam Count (BBC) Detector which covers the pseudorapidity range $-5.0<\eta<-3.3$. These results provide a direct comparison with previous measurements from PHENIX Collaboration with a similar centrality definition. We also employ the template fit method for nonflow subtraction and compare with results from the other three methods presented in the draft. The detailed simulation studies for the nonflow subtraction with HIJING is also presented.

\section{$v_n$ from BBC centrality}
\begin{figure}[h]
\centering
\includegraphics[width=0.6\linewidth]{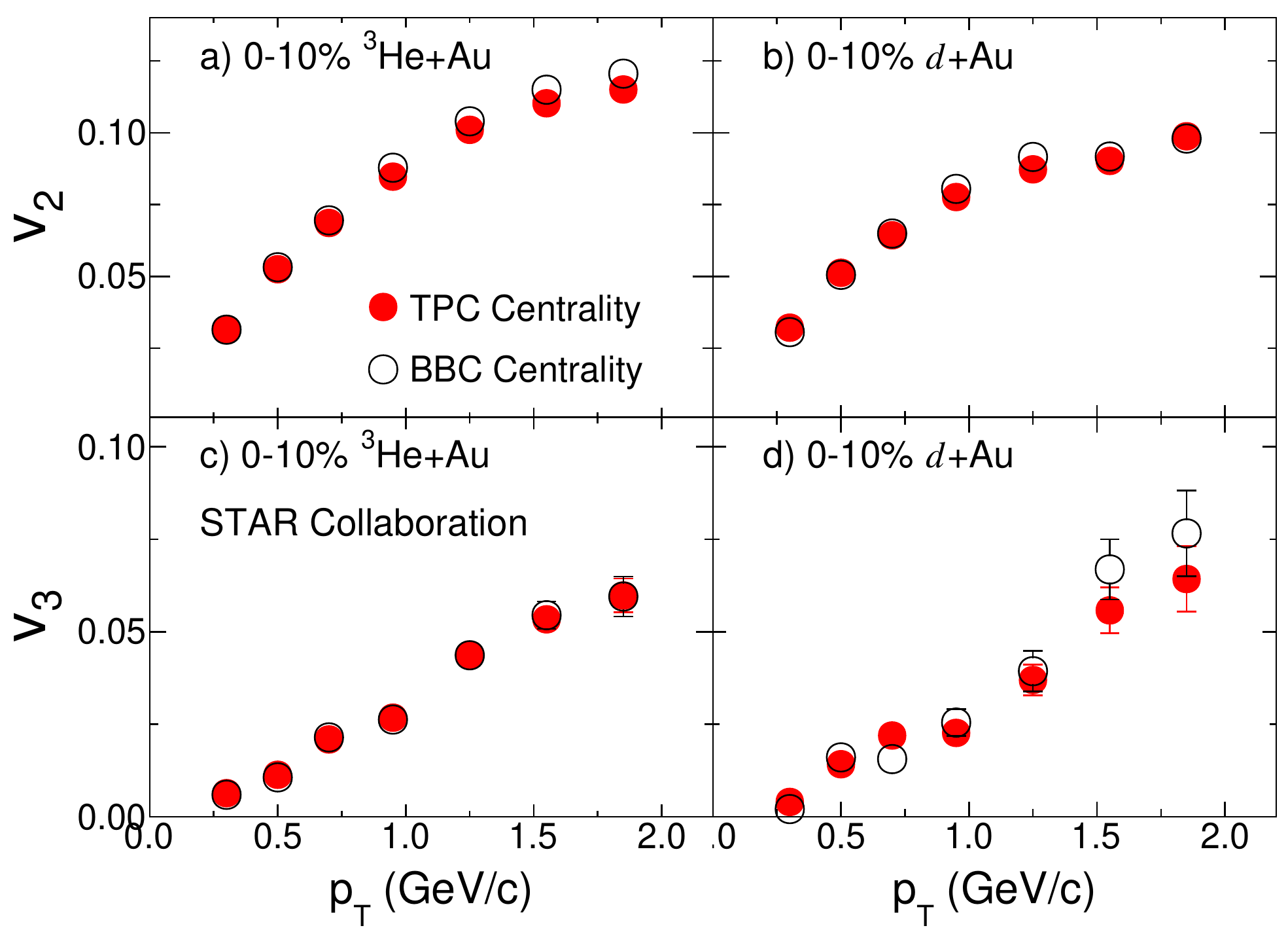}
\caption{The $v_n$ values obtained with TPC-centrality and BBC-centrality for 0-10\% \dau and \heau collisions.}\label{vnbbctpc}
\end{figure}

Two different centrality definitions are used to measure $v_{2,3}$ to check the impact from centrality definition. 
For the $d$+Au and $^{3}$He+Au MB data, the TPC and the BBC [on the Au-going side ($-5.0<\eta<-3.3$)] are used to 
select 0-10\% centrality events respectively. The $v_n$
values obtained with TPC - and BBC-centrality is shown in Fig.~\ref{vnbbctpc}. 
The results ultilize the $c_1$ nonflow subtraction method and are found to be consistent within statistical uncertainties.

\section{Template Fit}
The template-fit method is detailed in Ref.~\cite{Aad:2016}. In brief, the method assumes that the \Yphi\ distributions for $^{3}$He+Au, $d$+Au, and $p$+Au are superpositions of a scaled MB \Yphi\ distribution for \pp collisions [that characterizes the non-flow] and a constant modulated by the ridge $\sum_{n=2}^{4}\ c_{n}^{sub}\cos(n\dphi)$ as:
\begin{equation}
\YphiTempl = F\Yphipp +  \YphiRidge\, ,
\label{eq:template}
\end{equation}
where 
\begin{equation}
\YphiRidge = G \left(1 + 2\sum_{n=2}^{4}\ c_{n}^{sub} \cos{(n\Delta \phi)}\right)\, ,
\label{eq:template_ridge}
\end{equation}
%
with free parameters $F$ and $c_{n}^{sub}$. The coefficient $G$, which
represents the magnitude of the combinatorial component of $\YphiRidge$, is fixed by requiring
$\int_0^{\pi}{\mathrm{d}\dphi}\; Y^{\mathrm{templ}} = \int_0^{\pi}{\mathrm{d}\dphi} \;
Y^{\mathrm{HM}}$. 

Figure~\ref{fig:method} shows a comparison of the $v_{2,3}$ values extracted from template fit and comparison with other three nonflow subtraction methods. The comparison indicates the results from template fit is quite similar to that of method III and the difference are well within the systematic uncertainties signed for different subtraction methods.
\begin{figure}[h]
\begin{center}
  \includegraphics[width=1.0\linewidth]{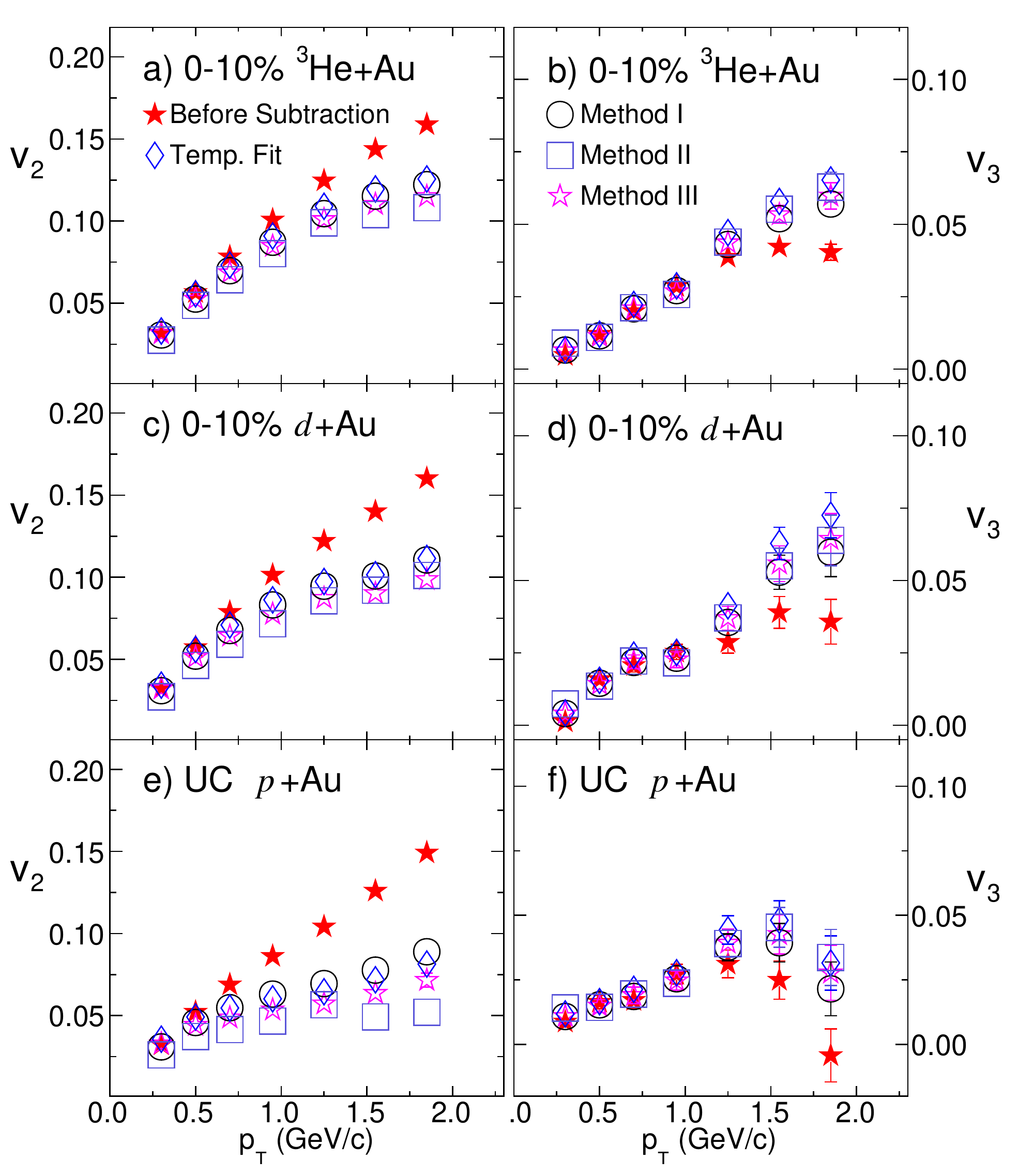}
  \caption{
Comparison of the flow coefficients $v_2$ and $v_3$, in $0\%-10\%$ $^{3}$He+Au, $0\%-10\%$ $d$+Au, and  $0\%-2\%$ $p$+Au collisions, before and after non-flow subtraction. The results for several methods of subtraction [discussed in the text] are presented as indicated.
The systematic uncertainties are not shown.
}
  \label{fig:method}
  \end{center}
\end{figure}

\section{Nonflow subtraction with HIJING simulation}
\label{hijing}

\begin{figure*}[ht]
\centering
\includegraphics[width=0.8\linewidth]{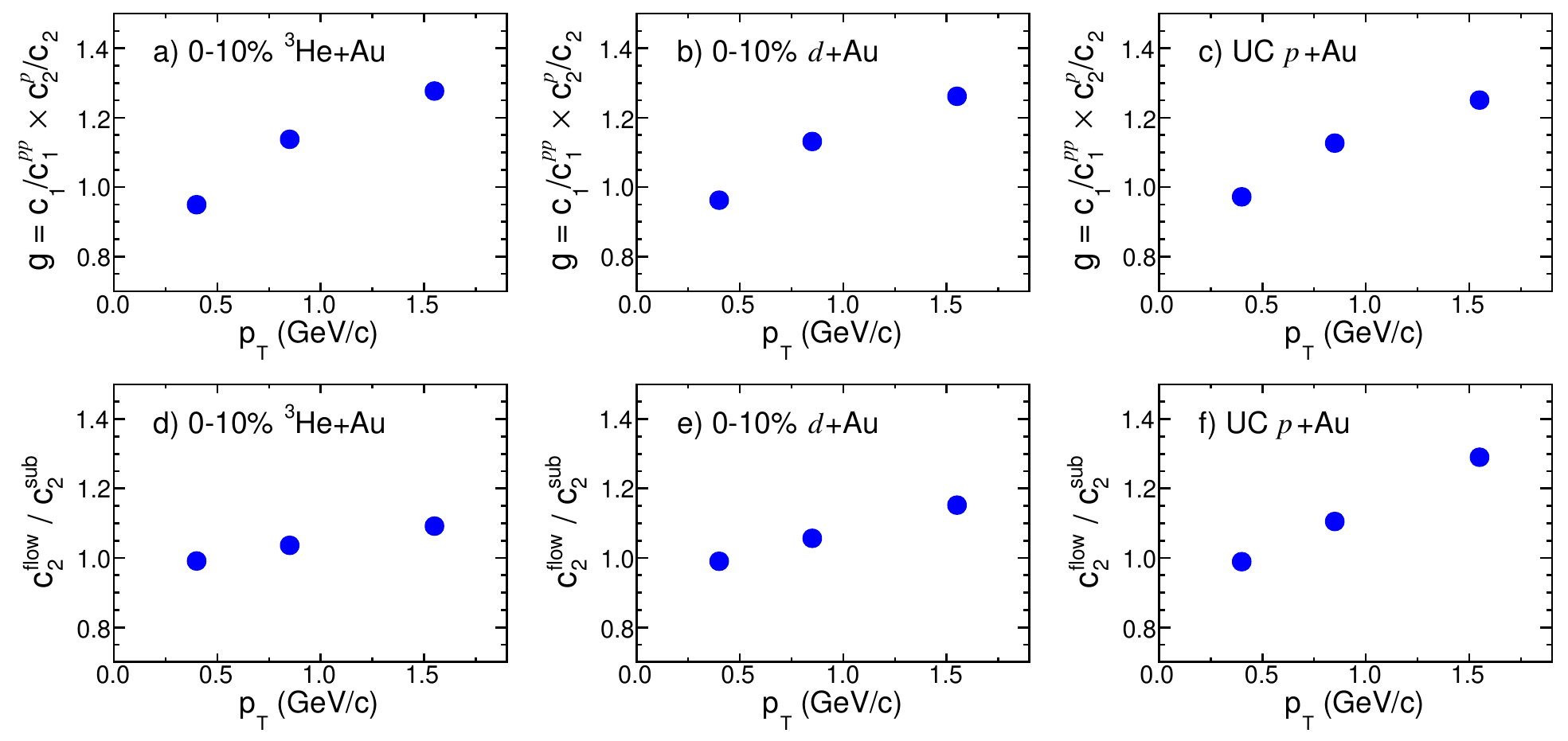}
\caption{The $g$ values and $c^{\mathrm{flow}}_{2}/c^{\mathrm{sub}}_{2}$ from HIJING as a function of $p_{\mathrm{T}}$ in 0-10\% \heau, 0-10\% \dau and UC \pau collisions.}\label{c2dAunonflow}
\end{figure*}

\begin{figure*}[]
\centering
\includegraphics[width=0.8\linewidth]{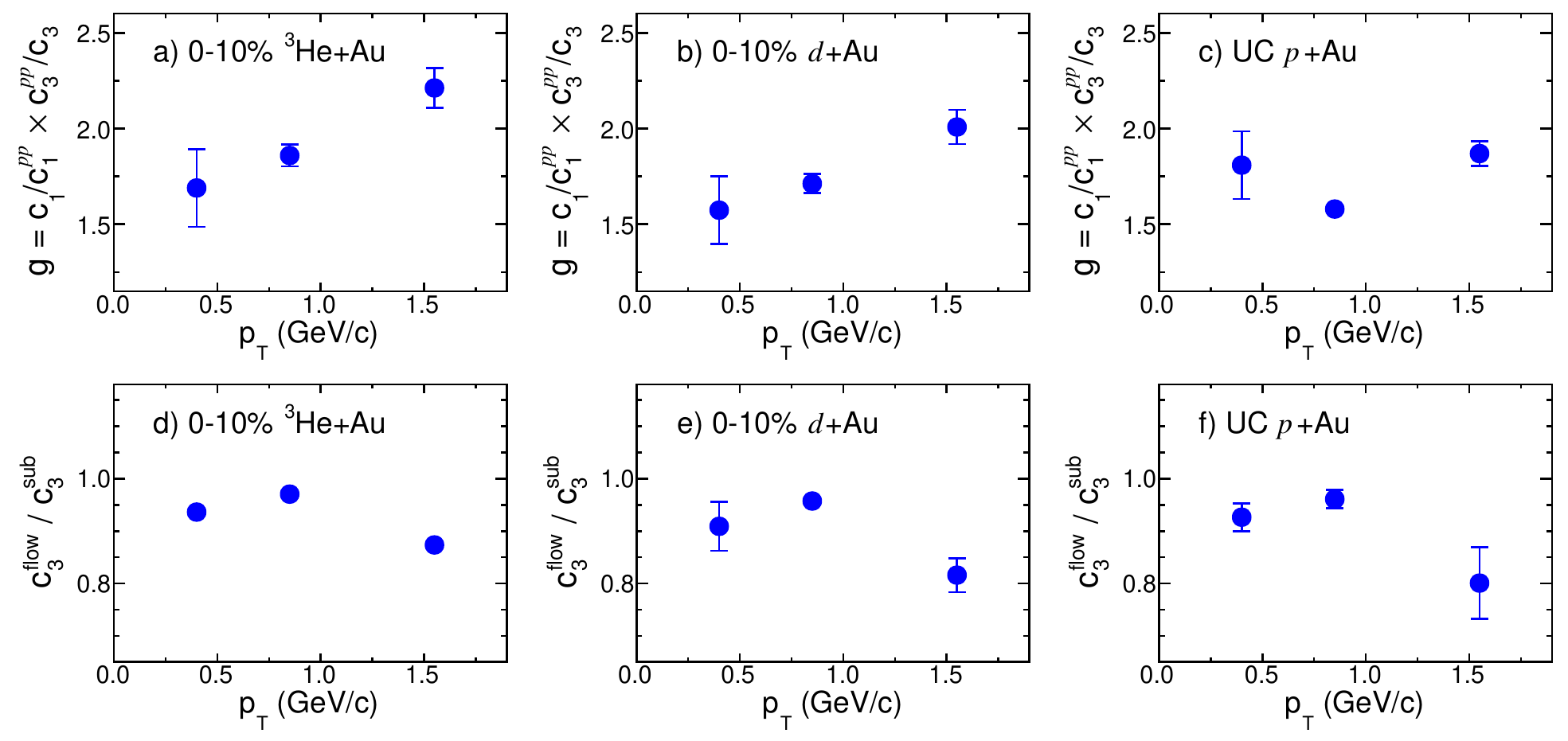}
\caption{The $g$ values and $c^{\mathrm{flow}}_{3}/c^{\mathrm{sub}}_{3}$ from HIJING as a function of $p_{\mathrm{T}}$ in 0-10\% \heau, 0-10\% \dau and UC \pau collisions.}\label{c3dAunonflow}
\end{figure*}

The nonflow contributions in UC \pau, 0-10\% \dau and 0-10\% \heau collisions are estimated by using $c_{n}$ from \pp collisions:
\begin{equation}
    c^{\mathrm{sub}}_{n} = c_{n} - f\times{c^{pp}_{n}}
    \label{nonflowequ}
\end{equation}
where $f$ is the ratio of $c_1$ between \pp and $p$/$d$/$^{3}$He+Au collisions for the $c_1$ subtraction method.

The true collective flow signal $c^{\mathrm{flow}}_{n}$ can be expressed as
\begin{equation}
    c^{\mathrm{flow}}_{n} = c_{n} - (f/g)\times{c^{pp}_{n}}
    \label{flowequ}
\end{equation}
where $g>$1 ($g<$1) means the nonflow is over(under)-estimated. 

Since $c^{\mathrm{flow}}_{n}=0$ in HIJING event generator, the value of $g$ can be extracted from Eq.~\ref{flowequ} as
\begin{equation}
    g = \frac{f\times{c^{pp}_{n}}}{c_{n}}. 
    \label{eq:g1}
\end{equation} 

\noindent The magnitude of over(under)-subtraction in real data can be estimated by 

\begin{equation}
    \frac{c^{\mathrm{flow}}_{n}}{c^{\mathrm{sub}}_{n}} = \frac{c_{n} -  (f/g)\times{c^{pp}_{n}}}{c_{n} - f\times{c^{pp}_{n}}}.
    \label{oversub}
\end{equation}

The $g$ and $c^{\mathrm{flow}}_{n}/c^{\mathrm{sub}}_{n}$ values are shown in Fig.~\ref{c2dAunonflow} and Fig.~\ref{c3dAunonflow} for $n=2$ and $n=3$ respectively. The overall uncertainties for nonflow subtraction are less than 25\% for $v_2$ and 20\%
for $v_3$ results, which is within the systematical uncertainties of the measurements.

\end{document}